\documentclass[journal]{IEEEtran}

\usepackage[utf8]{inputenc} 
\usepackage[T1]{fontenc}    
\usepackage{hyperref}       
\usepackage{url}            
\usepackage{booktabs}       
\usepackage{amsfonts}       
\usepackage{nicefrac}       
\usepackage{microtype}      
\usepackage{lipsum}
\usepackage{fancyhdr}       
\usepackage{graphicx}       
\graphicspath{{media/}}     
\usepackage{tabularx}
\usepackage{subcaption}

\usepackage{todonotes}
\usepackage[inline]{enumitem}

\usepackage{xcolor}

\usepackage{colortbl}
\PassOptionsToPackage{dvipsnames,table}{color}
\usepackage{pgfplotstable}
\pgfplotsset{compat=1.14}
\definecolor{b0}{rgb}{1.0000, 1.0000, 1.000}
 \definecolor{b1}{rgb}{0.990, 1.0000, 1.000}
 \definecolor{b2}{rgb}{0.980, 1.0000, 1.000}
 \definecolor{b3}{rgb}{0.970, 1.0000, 1.000}
 \definecolor{b4}{rgb}{0.960, 1.0000, 1.000}
 \definecolor{b5}{rgb}{0.950, 1.0000, 1.000}
 \definecolor{b6}{rgb}{0.940, 1.0000, 1.000}
 \definecolor{b7}{rgb}{0.930, 1.0000, 1.000}
 \definecolor{b8}{rgb}{0.920, 1.0000, 1.000}
 \definecolor{b9}{rgb}{0.910, 1.0000, 1.000}
\definecolor{b10}{rgb}{0.900, 1.0000, 1.000}
\definecolor{b11}{rgb}{0.890, 1.0000, 1.000}
\definecolor{b12}{rgb}{0.880, 1.0000, 1.000}
\definecolor{b13}{rgb}{0.870, 1.0000, 1.000}
\definecolor{b14}{rgb}{0.860, 1.0000, 1.000}
\definecolor{b15}{rgb}{0.850, 1.0000, 1.000}
\definecolor{b16}{rgb}{0.840, 1.0000, 1.000}
\definecolor{b17}{rgb}{0.830, 1.0000, 1.000}
\definecolor{b18}{rgb}{0.820, 1.0000, 1.000}
\definecolor{b19}{rgb}{0.810, 1.0000, 1.000}
\definecolor{b20}{rgb}{0.800, 1.0000, 1.000}
\definecolor{b21}{rgb}{0.790, 1.0000, 1.000}
\definecolor{b22}{rgb}{0.780, 1.0000, 1.000}
\definecolor{b23}{rgb}{0.770, 1.0000, 1.000}
\definecolor{b24}{rgb}{0.760, 1.0000, 1.000}
\definecolor{b25}{rgb}{0.750, 1.0000, 1.000}
\definecolor{b26}{rgb}{0.740, 1.0000, 1.000}
\definecolor{b27}{rgb}{0.730, 1.0000, 1.000}
\definecolor{b28}{rgb}{0.720, 1.0000, 1.000}
\definecolor{b29}{rgb}{0.710, 1.0000, 1.000}
\definecolor{b30}{rgb}{0.700, 1.0000, 1.000}
\definecolor{b31}{rgb}{0.690, 1.0000, 1.000}
\definecolor{b32}{rgb}{0.680, 1.0000, 1.000}
\definecolor{b33}{rgb}{0.670, 1.0000, 1.000}
\definecolor{b34}{rgb}{0.660, 1.0000, 1.000}
\definecolor{b35}{rgb}{0.650, 1.0000, 1.000}
\definecolor{b36}{rgb}{0.640, 1.0000, 1.000}
\definecolor{b37}{rgb}{0.630, 1.0000, 1.000}
\definecolor{b38}{rgb}{0.620, 1.0000, 1.000}
\definecolor{b39}{rgb}{0.610, 1.0000, 1.000}
\definecolor{b40}{rgb}{0.600, 1.0000, 1.000}
\definecolor{b41}{rgb}{0.590, 1.0000, 1.000}
\definecolor{b42}{rgb}{0.580, 1.0000, 1.000}
\definecolor{b43}{rgb}{0.570, 1.0000, 1.000}
\definecolor{b44}{rgb}{0.560, 1.0000, 1.000}
\definecolor{b45}{rgb}{0.550, 1.0000, 1.000}
\definecolor{b46}{rgb}{0.540, 1.0000, 1.000}
\definecolor{b47}{rgb}{0.530, 1.0000, 1.000}
\definecolor{b48}{rgb}{0.520, 1.0000, 1.000}
\definecolor{b49}{rgb}{0.510, 1.0000, 1.000}
\definecolor{b50}{rgb}{0.500, 1.0000, 1.000}
\definecolor{b51}{rgb}{0.490, 1.0000, 1.000}
\definecolor{b52}{rgb}{0.480, 1.0000, 1.000}
\definecolor{b53}{rgb}{0.470, 1.0000, 1.000}
\definecolor{b54}{rgb}{0.460, 1.0000, 1.000}
\definecolor{b55}{rgb}{0.450, 1.0000, 1.000}
\definecolor{b56}{rgb}{0.440, 1.0000, 1.000}
\definecolor{b57}{rgb}{0.430, 1.0000, 1.000}
\definecolor{b58}{rgb}{0.420, 1.0000, 1.000}
\definecolor{b59}{rgb}{0.410, 1.0000, 1.000}
\definecolor{b60}{rgb}{0.400, 1.0000, 1.000}
\definecolor{b61}{rgb}{0.390, 1.0000, 1.000}
\definecolor{b62}{rgb}{0.380, 1.0000, 1.000}
\definecolor{b63}{rgb}{0.370, 1.0000, 1.000}
\definecolor{b64}{rgb}{0.360, 1.0000, 1.000}
\definecolor{b65}{rgb}{0.350, 1.0000, 1.000}
\definecolor{b66}{rgb}{0.340, 1.0000, 1.000}
\definecolor{b67}{rgb}{0.330, 1.0000, 1.000}
\definecolor{b68}{rgb}{0.320, 1.0000, 1.000}
\definecolor{b69}{rgb}{0.310, 1.0000, 1.000}
\definecolor{b70}{rgb}{0.300, 1.0000, 1.000}
\definecolor{b71}{rgb}{0.290, 1.0000, 1.000}
\definecolor{b72}{rgb}{0.280, 1.0000, 1.000}
\definecolor{b73}{rgb}{0.270, 1.0000, 1.000}
\definecolor{b74}{rgb}{0.260, 1.0000, 1.000}
\definecolor{b75}{rgb}{0.250, 1.0000, 1.000}
\definecolor{b76}{rgb}{0.240, 1.0000, 1.000}
\definecolor{b77}{rgb}{0.230, 1.0000, 1.000}
\definecolor{b78}{rgb}{0.220, 1.0000, 1.000}
\definecolor{b79}{rgb}{0.210, 1.0000, 1.000}
\definecolor{b80}{rgb}{0.200, 1.0000, 1.000}
\definecolor{b81}{rgb}{0.190, 1.0000, 1.000}
\definecolor{b82}{rgb}{0.180, 1.0000, 1.000}
\definecolor{b83}{rgb}{0.170, 1.0000, 1.000}
\definecolor{b84}{rgb}{0.160, 1.0000, 1.000}
\definecolor{b85}{rgb}{0.150, 1.0000, 1.000}
\definecolor{b86}{rgb}{0.140, 1.0000, 1.000}
\definecolor{b87}{rgb}{0.130, 1.0000, 1.000}
\definecolor{b88}{rgb}{0.120, 1.0000, 1.000}
\definecolor{b89}{rgb}{0.110, 1.0000, 1.000}
\definecolor{b90}{rgb}{0.100, 1.0000, 1.000}
\definecolor{b91}{rgb}{0.090, 1.0000, 1.000}
\definecolor{b92}{rgb}{0.080, 1.0000, 1.000}
\definecolor{b93}{rgb}{0.070, 1.0000, 1.000}
\definecolor{b94}{rgb}{0.060, 1.0000, 1.000}
\definecolor{b95}{rgb}{0.050, 1.0000, 1.000}
\definecolor{b96}{rgb}{0.040, 1.0000, 1.000}
\definecolor{b97}{rgb}{0.030, 1.0000, 1.000}
\definecolor{b98}{rgb}{0.020, 1.0000, 1.000}
\definecolor{b99}{rgb}{0.010, 1.0000, 1.000}
\definecolor{b100}{rgb}{0.000, 1.0000, 1.000}

\definecolor{p0}{rgb}{1.000, 1.000, 1.000}
\definecolor{p1}{rgb}{1.000, 0.990, 1.000}
\definecolor{p2}{rgb}{1.000, 0.980, 1.000}
\definecolor{p3}{rgb}{1.000, 0.970, 1.000}
\definecolor{p4}{rgb}{1.000, 0.960, 1.000}
\definecolor{p5}{rgb}{1.000, 0.950, 1.000}
\definecolor{p6}{rgb}{1.000, 0.940, 1.000}
\definecolor{p7}{rgb}{1.000, 0.930, 1.000}
\definecolor{p8}{rgb}{1.000, 0.920, 1.000}
\definecolor{p9}{rgb}{1.000, 0.910, 1.000}
\definecolor{p10}{rgb}{1.000, 0.900, 1.000}
\definecolor{p11}{rgb}{1.000, 0.890, 1.000}
\definecolor{p12}{rgb}{1.000, 0.880, 1.000}
\definecolor{p13}{rgb}{1.000, 0.870, 1.000}
\definecolor{p14}{rgb}{1.000, 0.860, 1.000}
\definecolor{p15}{rgb}{1.000, 0.850, 1.000}
\definecolor{p16}{rgb}{1.000, 0.840, 1.000}
\definecolor{p17}{rgb}{1.000, 0.830, 1.000}
\definecolor{p18}{rgb}{1.000, 0.820, 1.000}
\definecolor{p19}{rgb}{1.000, 0.810, 1.000}
\definecolor{p20}{rgb}{1.000, 0.800, 1.000}
\definecolor{p21}{rgb}{1.000, 0.790, 1.000}
\definecolor{p22}{rgb}{1.000, 0.780, 1.000}
\definecolor{p23}{rgb}{1.000, 0.770, 1.000}
\definecolor{p24}{rgb}{1.000, 0.760, 1.000}
\definecolor{p25}{rgb}{1.000, 0.750, 1.000}
\definecolor{p26}{rgb}{1.000, 0.740, 1.000}
\definecolor{p27}{rgb}{1.000, 0.730, 1.000}
\definecolor{p28}{rgb}{1.000, 0.720, 1.000}
\definecolor{p29}{rgb}{1.000, 0.710, 1.000}
\definecolor{p30}{rgb}{1.000, 0.700, 1.000}
\definecolor{p31}{rgb}{1.000, 0.690, 1.000}
\definecolor{p32}{rgb}{1.000, 0.680, 1.000}
\definecolor{p33}{rgb}{1.000, 0.670, 1.000}
\definecolor{p34}{rgb}{1.000, 0.660, 1.000}
\definecolor{p35}{rgb}{1.000, 0.650, 1.000}
\definecolor{p36}{rgb}{1.000, 0.640, 1.000}
\definecolor{p37}{rgb}{1.000, 0.630, 1.000}
\definecolor{p38}{rgb}{1.000, 0.620, 1.000}
\definecolor{p39}{rgb}{1.000, 0.610, 1.000}
\definecolor{p40}{rgb}{1.000, 0.600, 1.000}
\definecolor{p41}{rgb}{1.000, 0.590, 1.000}
\definecolor{p42}{rgb}{1.000, 0.580, 1.000}
\definecolor{p43}{rgb}{1.000, 0.570, 1.000}
\definecolor{p44}{rgb}{1.000, 0.560, 1.000}
\definecolor{p45}{rgb}{1.000, 0.550, 1.000}
\definecolor{p46}{rgb}{1.000, 0.540, 1.000}
\definecolor{p47}{rgb}{1.000, 0.530, 1.000}
\definecolor{p48}{rgb}{1.000, 0.520, 1.000}
\definecolor{p49}{rgb}{1.000, 0.510, 1.000}
\definecolor{p50}{rgb}{1.000, 0.500, 1.000}
\definecolor{p51}{rgb}{1.000, 0.490, 1.000}
\definecolor{p52}{rgb}{1.000, 0.480, 1.000}
\definecolor{p53}{rgb}{1.000, 0.470, 1.000}
\definecolor{p54}{rgb}{1.000, 0.460, 1.000}
\definecolor{p55}{rgb}{1.000, 0.450, 1.000}
\definecolor{p56}{rgb}{1.000, 0.440, 1.000}
\definecolor{p57}{rgb}{1.000, 0.430, 1.000}
\definecolor{p58}{rgb}{1.000, 0.420, 1.000}
\definecolor{p59}{rgb}{1.000, 0.410, 1.000}
\definecolor{p60}{rgb}{1.000, 0.400, 1.000}
\definecolor{p61}{rgb}{1.000, 0.390, 1.000}
\definecolor{p62}{rgb}{1.000, 0.380, 1.000}
\definecolor{p63}{rgb}{1.000, 0.370, 1.000}
\definecolor{p64}{rgb}{1.000, 0.360, 1.000}
\definecolor{p65}{rgb}{1.000, 0.350, 1.000}
\definecolor{p66}{rgb}{1.000, 0.340, 1.000}
\definecolor{p67}{rgb}{1.000, 0.330, 1.000}
\definecolor{p68}{rgb}{1.000, 0.320, 1.000}
\definecolor{p69}{rgb}{1.000, 0.310, 1.000}
\definecolor{p70}{rgb}{1.000, 0.300, 1.000}
\definecolor{p71}{rgb}{1.000, 0.290, 1.000}
\definecolor{p72}{rgb}{1.000, 0.280, 1.000}
\definecolor{p73}{rgb}{1.000, 0.270, 1.000}
\definecolor{p74}{rgb}{1.000, 0.260, 1.000}
\definecolor{p75}{rgb}{1.000, 0.250, 1.000}
\definecolor{p76}{rgb}{1.000, 0.240, 1.000}
\definecolor{p77}{rgb}{1.000, 0.230, 1.000}
\definecolor{p78}{rgb}{1.000, 0.220, 1.000}
\definecolor{p79}{rgb}{1.000, 0.210, 1.000}
\definecolor{p80}{rgb}{1.000, 0.200, 1.000}
\definecolor{p81}{rgb}{1.000, 0.190, 1.000}
\definecolor{p82}{rgb}{1.000, 0.180, 1.000}
\definecolor{p83}{rgb}{1.000, 0.170, 1.000}
\definecolor{p84}{rgb}{1.000, 0.160, 1.000}
\definecolor{p85}{rgb}{1.000, 0.150, 1.000}
\definecolor{p86}{rgb}{1.000, 0.140, 1.000}
\definecolor{p87}{rgb}{1.000, 0.130, 1.000}
\definecolor{p88}{rgb}{1.000, 0.120, 1.000}
\definecolor{p89}{rgb}{1.000, 0.110, 1.000}
\definecolor{p90}{rgb}{1.000, 0.100, 1.000}
\definecolor{p91}{rgb}{1.000, 0.090, 1.000}
\definecolor{p92}{rgb}{1.000, 0.080, 1.000}
\definecolor{p93}{rgb}{1.000, 0.070, 1.000}
\definecolor{p94}{rgb}{1.000, 0.060, 1.000}
\definecolor{p95}{rgb}{1.000, 0.050, 1.000}
\definecolor{p96}{rgb}{1.000, 0.040, 1.000}
\definecolor{p97}{rgb}{1.000, 0.030, 1.000}
\definecolor{p98}{rgb}{1.000, 0.020, 1.000}
\definecolor{p99}{rgb}{1.000, 0.010, 1.000}
\definecolor{p100}{rgb}{1.000, 0.000, 1.000}

\definecolor{v0}{rgb}{1.000, 1.000, 1.000}
 \definecolor{v1}{rgb}{0.990, 0.990, 1.000}
 \definecolor{v2}{rgb}{0.980, 0.980, 1.000}
 \definecolor{v3}{rgb}{0.970, 0.970, 1.000}
 \definecolor{v4}{rgb}{0.960, 0.960, 1.000}
 \definecolor{v5}{rgb}{0.950, 0.950, 1.000}
 \definecolor{v6}{rgb}{0.940, 0.940, 1.000}
 \definecolor{v7}{rgb}{0.930, 0.930, 1.000}
 \definecolor{v8}{rgb}{0.920, 0.920, 1.000}
 \definecolor{v9}{rgb}{0.910, 0.910, 1.000}
\definecolor{v10}{rgb}{0.900, 0.900, 1.000}
\definecolor{v11}{rgb}{0.890, 0.890, 1.000}
\definecolor{v12}{rgb}{0.880, 0.880, 1.000}
\definecolor{v13}{rgb}{0.870, 0.870, 1.000}
\definecolor{v14}{rgb}{0.860, 0.860, 1.000}
\definecolor{v15}{rgb}{0.850, 0.850, 1.000}
\definecolor{v16}{rgb}{0.840, 0.840, 1.000}
\definecolor{v17}{rgb}{0.830, 0.830, 1.000}
\definecolor{v18}{rgb}{0.820, 0.820, 1.000}
\definecolor{v19}{rgb}{0.810, 0.810, 1.000}
\definecolor{v20}{rgb}{0.800, 0.800, 1.000}
\definecolor{v21}{rgb}{0.790, 0.790, 1.000}
\definecolor{v22}{rgb}{0.780, 0.780, 1.000}
\definecolor{v23}{rgb}{0.770, 0.770, 1.000}
\definecolor{v24}{rgb}{0.760, 0.760, 1.000}
\definecolor{v25}{rgb}{0.750, 0.750, 1.000}
\definecolor{v26}{rgb}{0.740, 0.740, 1.000}
\definecolor{v27}{rgb}{0.730, 0.730, 1.000}
\definecolor{v28}{rgb}{0.720, 0.720, 1.000}
\definecolor{v29}{rgb}{0.710, 0.710, 1.000}
\definecolor{v30}{rgb}{0.700, 0.700, 1.000}
\definecolor{v31}{rgb}{0.690, 0.690, 1.000}
\definecolor{v32}{rgb}{0.680, 0.680, 1.000}
\definecolor{v33}{rgb}{0.670, 0.670, 1.000}
\definecolor{v34}{rgb}{0.660, 0.660, 1.000}
\definecolor{v35}{rgb}{0.650, 0.650, 1.000}
\definecolor{v36}{rgb}{0.640, 0.640, 1.000}
\definecolor{v37}{rgb}{0.630, 0.630, 1.000}
\definecolor{v38}{rgb}{0.620, 0.620, 1.000}
\definecolor{v39}{rgb}{0.610, 0.610, 1.000}
\definecolor{v40}{rgb}{0.600, 0.600, 1.000}
\definecolor{v41}{rgb}{0.590, 0.590, 1.000}
\definecolor{v42}{rgb}{0.580, 0.580, 1.000}
\definecolor{v43}{rgb}{0.570, 0.570, 1.000}
\definecolor{v44}{rgb}{0.560, 0.560, 1.000}
\definecolor{v45}{rgb}{0.550, 0.550, 1.000}
\definecolor{v46}{rgb}{0.540, 0.540, 1.000}
\definecolor{v47}{rgb}{0.530, 0.530, 1.000}
\definecolor{v48}{rgb}{0.520, 0.520, 1.000}
\definecolor{v49}{rgb}{0.510, 0.510, 1.000}
\definecolor{v50}{rgb}{0.500, 0.500, 1.000}
\definecolor{v51}{rgb}{0.490, 0.490, 1.000}
\definecolor{v52}{rgb}{0.480, 0.480, 1.000}
\definecolor{v53}{rgb}{0.470, 0.470, 1.000}
\definecolor{v54}{rgb}{0.460, 0.460, 1.000}
\definecolor{v55}{rgb}{0.450, 0.450, 1.000}
\definecolor{v56}{rgb}{0.440, 0.440, 1.000}
\definecolor{v57}{rgb}{0.430, 0.430, 1.000}
\definecolor{v58}{rgb}{0.420, 0.420, 1.000}
\definecolor{v59}{rgb}{0.410, 0.410, 1.000}
\definecolor{v60}{rgb}{0.400, 0.400, 1.000}
\definecolor{v61}{rgb}{0.390, 0.390, 1.000}
\definecolor{v62}{rgb}{0.380, 0.380, 1.000}
\definecolor{v63}{rgb}{0.370, 0.370, 1.000}
\definecolor{v64}{rgb}{0.360, 0.360, 1.000}
\definecolor{v65}{rgb}{0.350, 0.350, 1.000}
\definecolor{v66}{rgb}{0.340, 0.340, 1.000}
\definecolor{v67}{rgb}{0.330, 0.330, 1.000}
\definecolor{v68}{rgb}{0.320, 0.320, 1.000}
\definecolor{v69}{rgb}{0.310, 0.310, 1.000}
\definecolor{v70}{rgb}{0.300, 0.300, 1.000}
\definecolor{v71}{rgb}{0.290, 0.290, 1.000}
\definecolor{v72}{rgb}{0.280, 0.280, 1.000}
\definecolor{v73}{rgb}{0.270, 0.270, 1.000}
\definecolor{v74}{rgb}{0.260, 0.260, 1.000}
\definecolor{v75}{rgb}{0.250, 0.250, 1.000}
\definecolor{v76}{rgb}{0.240, 0.240, 1.000}
\definecolor{v77}{rgb}{0.230, 0.230, 1.000}
\definecolor{v78}{rgb}{0.220, 0.220, 1.000}
\definecolor{v79}{rgb}{0.210, 0.210, 1.000}
\definecolor{v80}{rgb}{0.200, 0.200, 1.000}
\definecolor{v81}{rgb}{0.190, 0.190, 1.000}
\definecolor{v82}{rgb}{0.180, 0.180, 1.000}
\definecolor{v83}{rgb}{0.170, 0.170, 1.000}
\definecolor{v84}{rgb}{0.160, 0.160, 1.000}
\definecolor{v85}{rgb}{0.150, 0.150, 1.000}
\definecolor{v86}{rgb}{0.140, 0.140, 1.000}
\definecolor{v87}{rgb}{0.130, 0.130, 1.000}
\definecolor{v88}{rgb}{0.120, 0.120, 1.000}
\definecolor{v89}{rgb}{0.110, 0.110, 1.000}
\definecolor{v90}{rgb}{0.100, 0.100, 1.000}
\definecolor{v91}{rgb}{0.090, 0.090, 1.000}
\definecolor{v92}{rgb}{0.080, 0.080, 1.000}
\definecolor{v93}{rgb}{0.070, 0.070, 1.000}
\definecolor{v94}{rgb}{0.060, 0.060, 1.000}
\definecolor{v95}{rgb}{0.050, 0.050, 1.000}
\definecolor{v96}{rgb}{0.040, 0.040, 1.000}
\definecolor{v97}{rgb}{0.030, 0.030, 1.000}
\definecolor{v98}{rgb}{0.020, 0.020, 1.000}
\definecolor{v99}{rgb}{0.010, 0.010, 1.000}
\definecolor{v100}{rgb}{0.000, 0.000, 1.000}

\definecolor{g0}{rgb}{1.000, 1.000, 1.000}
\definecolor{g1}{rgb}{0.990, 1.000, 0.990}
\definecolor{g2}{rgb}{0.980, 1.000, 0.980}
\definecolor{g3}{rgb}{0.970, 1.000, 0.970}
\definecolor{g4}{rgb}{0.960, 1.000, 0.960}
\definecolor{g5}{rgb}{0.950, 1.000, 0.950}
\definecolor{g6}{rgb}{0.940, 1.000, 0.940}
\definecolor{g7}{rgb}{0.930, 1.000, 0.930}
\definecolor{g8}{rgb}{0.920, 1.000, 0.920}
\definecolor{g9}{rgb}{0.910, 1.000, 0.910}
\definecolor{g10}{rgb}{0.900, 1.000, 0.900}
\definecolor{g11}{rgb}{0.890, 1.000, 0.890}
\definecolor{g12}{rgb}{0.880, 1.000, 0.880}
\definecolor{g13}{rgb}{0.870, 1.000, 0.870}
\definecolor{g14}{rgb}{0.860, 1.000, 0.860}
\definecolor{g15}{rgb}{0.850, 1.000, 0.850}
\definecolor{g16}{rgb}{0.840, 1.000, 0.840}
\definecolor{g17}{rgb}{0.830, 1.000, 0.830}
\definecolor{g18}{rgb}{0.820, 1.000, 0.820}
\definecolor{g19}{rgb}{0.810, 1.000, 0.810}
\definecolor{g20}{rgb}{0.800, 1.000, 0.800}
\definecolor{g21}{rgb}{0.790, 1.000, 0.790}
\definecolor{g22}{rgb}{0.780, 1.000, 0.780}
\definecolor{g23}{rgb}{0.770, 1.000, 0.770}
\definecolor{g24}{rgb}{0.760, 1.000, 0.760}
\definecolor{g25}{rgb}{0.750, 1.000, 0.750}
\definecolor{g26}{rgb}{0.740, 1.000, 0.740}
\definecolor{g27}{rgb}{0.730, 1.000, 0.730}
\definecolor{g28}{rgb}{0.720, 1.000, 0.720}
\definecolor{g29}{rgb}{0.710, 1.000, 0.710}
\definecolor{g30}{rgb}{0.700, 1.000, 0.700}
\definecolor{g31}{rgb}{0.690, 1.000, 0.690}
\definecolor{g32}{rgb}{0.680, 1.000, 0.680}
\definecolor{g33}{rgb}{0.670, 1.000, 0.670}
\definecolor{g34}{rgb}{0.660, 1.000, 0.660}
\definecolor{g35}{rgb}{0.650, 1.000, 0.650}
\definecolor{g36}{rgb}{0.640, 1.000, 0.640}
\definecolor{g37}{rgb}{0.630, 1.000, 0.630}
\definecolor{g38}{rgb}{0.620, 1.000, 0.620}
\definecolor{g39}{rgb}{0.610, 1.000, 0.610}
\definecolor{g40}{rgb}{0.600, 1.000, 0.600}
\definecolor{g41}{rgb}{0.590, 1.000, 0.590}
\definecolor{g42}{rgb}{0.580, 1.000, 0.580}
\definecolor{g43}{rgb}{0.570, 1.000, 0.570}
\definecolor{g44}{rgb}{0.560, 1.000, 0.560}
\definecolor{g45}{rgb}{0.550, 1.000, 0.550}
\definecolor{g46}{rgb}{0.540, 1.000, 0.540}
\definecolor{g47}{rgb}{0.530, 1.000, 0.530}
\definecolor{g48}{rgb}{0.520, 1.000, 0.520}
\definecolor{g49}{rgb}{0.510, 1.000, 0.510}
\definecolor{g50}{rgb}{0.500, 1.000, 0.500}
\definecolor{g51}{rgb}{0.490, 1.000, 0.490}
\definecolor{g52}{rgb}{0.480, 1.000, 0.480}
\definecolor{g53}{rgb}{0.470, 1.000, 0.470}
\definecolor{g54}{rgb}{0.460, 1.000, 0.460}
\definecolor{g55}{rgb}{0.450, 1.000, 0.450}
\definecolor{g56}{rgb}{0.440, 1.000, 0.440}
\definecolor{g57}{rgb}{0.430, 1.000, 0.430}
\definecolor{g58}{rgb}{0.420, 1.000, 0.420}
\definecolor{g59}{rgb}{0.410, 1.000, 0.410}
\definecolor{g60}{rgb}{0.400, 1.000, 0.400}
\definecolor{g61}{rgb}{0.390, 1.000, 0.390}
\definecolor{g62}{rgb}{0.380, 1.000, 0.380}
\definecolor{g63}{rgb}{0.370, 1.000, 0.370}
\definecolor{g64}{rgb}{0.360, 1.000, 0.360}
\definecolor{g65}{rgb}{0.350, 1.000, 0.350}
\definecolor{g66}{rgb}{0.340, 1.000, 0.340}
\definecolor{g67}{rgb}{0.330, 1.000, 0.330}
\definecolor{g68}{rgb}{0.320, 1.000, 0.320}
\definecolor{g69}{rgb}{0.310, 1.000, 0.310}
\definecolor{g70}{rgb}{0.300, 1.000, 0.300}
\definecolor{g71}{rgb}{0.290, 1.000, 0.290}
\definecolor{g72}{rgb}{0.280, 1.000, 0.280}
\definecolor{g73}{rgb}{0.270, 1.000, 0.270}
\definecolor{g74}{rgb}{0.260, 1.000, 0.260}
\definecolor{g75}{rgb}{0.250, 1.000, 0.250}
\definecolor{g76}{rgb}{0.240, 1.000, 0.240}
\definecolor{g77}{rgb}{0.230, 1.000, 0.230}
\definecolor{g78}{rgb}{0.220, 1.000, 0.220}
\definecolor{g79}{rgb}{0.210, 1.000, 0.210}
\definecolor{g80}{rgb}{0.200, 1.000, 0.200}
\definecolor{g81}{rgb}{0.190, 1.000, 0.190}
\definecolor{g82}{rgb}{0.180, 1.000, 0.180}
\definecolor{g83}{rgb}{0.170, 1.000, 0.170}
\definecolor{g84}{rgb}{0.160, 1.000, 0.160}
\definecolor{g85}{rgb}{0.150, 1.000, 0.150}
\definecolor{g86}{rgb}{0.140, 1.000, 0.140}
\definecolor{g87}{rgb}{0.130, 1.000, 0.130}
\definecolor{g88}{rgb}{0.120, 1.000, 0.120}
\definecolor{g89}{rgb}{0.110, 1.000, 0.110}
\definecolor{g90}{rgb}{0.100, 1.000, 0.100}
\definecolor{g91}{rgb}{0.090, 1.000, 0.090}
\definecolor{g92}{rgb}{0.080, 1.000, 0.080}
\definecolor{g93}{rgb}{0.070, 1.000, 0.070}
\definecolor{g94}{rgb}{0.060, 1.000, 0.060}
\definecolor{g95}{rgb}{0.050, 1.000, 0.050}
\definecolor{g96}{rgb}{0.040, 1.000, 0.040}
\definecolor{g97}{rgb}{0.030, 1.000, 0.030}
\definecolor{g98}{rgb}{0.020, 1.000, 0.020}
\definecolor{g99}{rgb}{0.010, 1.000, 0.010}
\definecolor{g100}{rgb}{0.000, 1.000, 0.000}

\definecolor{r0}{rgb}{1.000, 1.000, 1.000}
\definecolor{r1}{rgb}{1.000, 0.990, 0.990}
\definecolor{r2}{rgb}{1.000, 0.980, 0.980}
\definecolor{r3}{rgb}{1.000, 0.970, 0.970}
\definecolor{r4}{rgb}{1.000, 0.960, 0.960}
\definecolor{r5}{rgb}{1.000, 0.950, 0.950}
\definecolor{r6}{rgb}{1.000, 0.940, 0.940}
\definecolor{r7}{rgb}{1.000, 0.930, 0.930}
\definecolor{r8}{rgb}{1.000, 0.920, 0.920}
\definecolor{r9}{rgb}{1.000, 0.910, 0.910}
\definecolor{r10}{rgb}{1.000, 0.900, 0.900}
\definecolor{r11}{rgb}{1.000, 0.890, 0.890}
\definecolor{r12}{rgb}{1.000, 0.880, 0.880}
\definecolor{r13}{rgb}{1.000, 0.870, 0.870}
\definecolor{r14}{rgb}{1.000, 0.860, 0.860}
\definecolor{r15}{rgb}{1.000, 0.850, 0.850}
\definecolor{r16}{rgb}{1.000, 0.840, 0.840}
\definecolor{r17}{rgb}{1.000, 0.830, 0.830}
\definecolor{r18}{rgb}{1.000, 0.820, 0.820}
\definecolor{r19}{rgb}{1.000, 0.810, 0.810}
\definecolor{r20}{rgb}{1.000, 0.800, 0.800}
\definecolor{r21}{rgb}{1.000, 0.790, 0.790}
\definecolor{r22}{rgb}{1.000, 0.780, 0.780}
\definecolor{r23}{rgb}{1.000, 0.770, 0.770}
\definecolor{r24}{rgb}{1.000, 0.760, 0.760}
\definecolor{r25}{rgb}{1.000, 0.750, 0.750}
\definecolor{r26}{rgb}{1.000, 0.740, 0.740}
\definecolor{r27}{rgb}{1.000, 0.730, 0.730}
\definecolor{r28}{rgb}{1.000, 0.720, 0.720}
\definecolor{r29}{rgb}{1.000, 0.710, 0.710}
\definecolor{r30}{rgb}{1.000, 0.700, 0.700}
\definecolor{r31}{rgb}{1.000, 0.690, 0.690}
\definecolor{r32}{rgb}{1.000, 0.680, 0.680}
\definecolor{r33}{rgb}{1.000, 0.670, 0.670}
\definecolor{r34}{rgb}{1.000, 0.660, 0.660}
\definecolor{r35}{rgb}{1.000, 0.650, 0.650}
\definecolor{r36}{rgb}{1.000, 0.640, 0.640}
\definecolor{r37}{rgb}{1.000, 0.630, 0.630}
\definecolor{r38}{rgb}{1.000, 0.620, 0.620}
\definecolor{r39}{rgb}{1.000, 0.610, 0.610}
\definecolor{r40}{rgb}{1.000, 0.600, 0.600}
\definecolor{r41}{rgb}{1.000, 0.590, 0.590}
\definecolor{r42}{rgb}{1.000, 0.580, 0.580}
\definecolor{r43}{rgb}{1.000, 0.570, 0.570}
\definecolor{r44}{rgb}{1.000, 0.560, 0.560}
\definecolor{r45}{rgb}{1.000, 0.550, 0.550}
\definecolor{r46}{rgb}{1.000, 0.540, 0.540}
\definecolor{r47}{rgb}{1.000, 0.530, 0.530}
\definecolor{r48}{rgb}{1.000, 0.520, 0.520}
\definecolor{r49}{rgb}{1.000, 0.510, 0.510}
\definecolor{r50}{rgb}{1.000, 0.500, 0.500}
\definecolor{r51}{rgb}{1.000, 0.490, 0.490}
\definecolor{r52}{rgb}{1.000, 0.480, 0.480}
\definecolor{r53}{rgb}{1.000, 0.470, 0.470}
\definecolor{r54}{rgb}{1.000, 0.460, 0.460}
\definecolor{r55}{rgb}{1.000, 0.450, 0.450}
\definecolor{r56}{rgb}{1.000, 0.440, 0.440}
\definecolor{r57}{rgb}{1.000, 0.430, 0.430}
\definecolor{r58}{rgb}{1.000, 0.420, 0.420}
\definecolor{r59}{rgb}{1.000, 0.410, 0.410}
\definecolor{r60}{rgb}{1.000, 0.400, 0.400}
\definecolor{r61}{rgb}{1.000, 0.390, 0.390}
\definecolor{r62}{rgb}{1.000, 0.380, 0.380}
\definecolor{r63}{rgb}{1.000, 0.370, 0.370}
\definecolor{r64}{rgb}{1.000, 0.360, 0.360}
\definecolor{r65}{rgb}{1.000, 0.350, 0.350}
\definecolor{r66}{rgb}{1.000, 0.340, 0.340}
\definecolor{r67}{rgb}{1.000, 0.330, 0.330}
\definecolor{r68}{rgb}{1.000, 0.320, 0.320}
\definecolor{r69}{rgb}{1.000, 0.310, 0.310}
\definecolor{r70}{rgb}{1.000, 0.300, 0.300}
\definecolor{r71}{rgb}{1.000, 0.290, 0.290}
\definecolor{r72}{rgb}{1.000, 0.280, 0.280}
\definecolor{r73}{rgb}{1.000, 0.270, 0.270}
\definecolor{r74}{rgb}{1.000, 0.260, 0.260}
\definecolor{r75}{rgb}{1.000, 0.250, 0.250}
\definecolor{r76}{rgb}{1.000, 0.240, 0.240}
\definecolor{r77}{rgb}{1.000, 0.230, 0.230}
\definecolor{r78}{rgb}{1.000, 0.220, 0.220}
\definecolor{r79}{rgb}{1.000, 0.210, 0.210}
\definecolor{r80}{rgb}{1.000, 0.200, 0.200}
\definecolor{r81}{rgb}{1.000, 0.190, 0.190}
\definecolor{r82}{rgb}{1.000, 0.180, 0.180}
\definecolor{r83}{rgb}{1.000, 0.170, 0.170}
\definecolor{r84}{rgb}{1.000, 0.160, 0.160}
\definecolor{r85}{rgb}{1.000, 0.150, 0.150}
\definecolor{r86}{rgb}{1.000, 0.140, 0.140}
\definecolor{r87}{rgb}{1.000, 0.130, 0.130}
\definecolor{r88}{rgb}{1.000, 0.120, 0.120}
\definecolor{r89}{rgb}{1.000, 0.110, 0.110}
\definecolor{r90}{rgb}{1.000, 0.100, 0.100}
\definecolor{r91}{rgb}{1.000, 0.090, 0.090}
\definecolor{r92}{rgb}{1.000, 0.080, 0.080}
\definecolor{r93}{rgb}{1.000, 0.070, 0.070}
\definecolor{r94}{rgb}{1.000, 0.060, 0.060}
\definecolor{r95}{rgb}{1.000, 0.050, 0.050}
\definecolor{r96}{rgb}{1.000, 0.040, 0.040}
\definecolor{r97}{rgb}{1.000, 0.030, 0.030}
\definecolor{r98}{rgb}{1.000, 0.020, 0.020}
\definecolor{r99}{rgb}{1.000, 0.010, 0.010}
\definecolor{r100}{rgb}{1.000, 0.000, 0.000}

\definecolor{y0}{rgb}{1.000, 1.000, 1.000}
\definecolor{y1}{rgb}{1.000, 1.000, 0.990}
\definecolor{y2}{rgb}{1.000, 1.000, 0.980}
\definecolor{y3}{rgb}{1.000, 1.000, 0.970}
\definecolor{y4}{rgb}{1.000, 1.000, 0.960}
\definecolor{y5}{rgb}{1.000, 1.000, 0.950}
\definecolor{y6}{rgb}{1.000, 1.000, 0.940}
\definecolor{y7}{rgb}{1.000, 1.000, 0.930}
\definecolor{y8}{rgb}{1.000, 1.000, 0.920}
\definecolor{y9}{rgb}{1.000, 1.000, 0.910}
\definecolor{y10}{rgb}{1.000, 1.000, 0.900}
\definecolor{y11}{rgb}{1.000, 1.000, 0.890}
\definecolor{y12}{rgb}{1.000, 1.000, 0.880}
\definecolor{y13}{rgb}{1.000, 1.000, 0.870}
\definecolor{y14}{rgb}{1.000, 1.000, 0.860}
\definecolor{y15}{rgb}{1.000, 1.000, 0.850}
\definecolor{y16}{rgb}{1.000, 1.000, 0.840}
\definecolor{y17}{rgb}{1.000, 1.000, 0.830}
\definecolor{y18}{rgb}{1.000, 1.000, 0.820}
\definecolor{y19}{rgb}{1.000, 1.000, 0.810}
\definecolor{y20}{rgb}{1.000, 1.000, 0.800}
\definecolor{y21}{rgb}{1.000, 1.000, 0.790}
\definecolor{y22}{rgb}{1.000, 1.000, 0.780}
\definecolor{y23}{rgb}{1.000, 1.000, 0.770}
\definecolor{y24}{rgb}{1.000, 1.000, 0.760}
\definecolor{y25}{rgb}{1.000, 1.000, 0.750}
\definecolor{y26}{rgb}{1.000, 1.000, 0.740}
\definecolor{y27}{rgb}{1.000, 1.000, 0.730}
\definecolor{y28}{rgb}{1.000, 1.000, 0.720}
\definecolor{y29}{rgb}{1.000, 1.000, 0.710}
\definecolor{y30}{rgb}{1.000, 1.000, 0.700}
\definecolor{y31}{rgb}{1.000, 1.000, 0.690}
\definecolor{y32}{rgb}{1.000, 1.000, 0.680}
\definecolor{y33}{rgb}{1.000, 1.000, 0.670}
\definecolor{y34}{rgb}{1.000, 1.000, 0.660}
\definecolor{y35}{rgb}{1.000, 1.000, 0.650}
\definecolor{y36}{rgb}{1.000, 1.000, 0.640}
\definecolor{y37}{rgb}{1.000, 1.000, 0.630}
\definecolor{y38}{rgb}{1.000, 1.000, 0.620}
\definecolor{y39}{rgb}{1.000, 1.000, 0.610}
\definecolor{y40}{rgb}{1.000, 1.000, 0.600}
\definecolor{y41}{rgb}{1.000, 1.000, 0.590}
\definecolor{y42}{rgb}{1.000, 1.000, 0.580}
\definecolor{y43}{rgb}{1.000, 1.000, 0.570}
\definecolor{y44}{rgb}{1.000, 1.000, 0.560}
\definecolor{y45}{rgb}{1.000, 1.000, 0.550}
\definecolor{y46}{rgb}{1.000, 1.000, 0.540}
\definecolor{y47}{rgb}{1.000, 1.000, 0.530}
\definecolor{y48}{rgb}{1.000, 1.000, 0.520}
\definecolor{y49}{rgb}{1.000, 1.000, 0.510}
\definecolor{y50}{rgb}{1.000, 1.000, 0.500}
\definecolor{y51}{rgb}{1.000, 1.000, 0.490}
\definecolor{y52}{rgb}{1.000, 1.000, 0.480}
\definecolor{y53}{rgb}{1.000, 1.000, 0.470}
\definecolor{y54}{rgb}{1.000, 1.000, 0.460}
\definecolor{y55}{rgb}{1.000, 1.000, 0.450}
\definecolor{y56}{rgb}{1.000, 1.000, 0.440}
\definecolor{y57}{rgb}{1.000, 1.000, 0.430}
\definecolor{y58}{rgb}{1.000, 1.000, 0.420}
\definecolor{y59}{rgb}{1.000, 1.000, 0.410}
\definecolor{y60}{rgb}{1.000, 1.000, 0.400}
\definecolor{y61}{rgb}{1.000, 1.000, 0.390}
\definecolor{y62}{rgb}{1.000, 1.000, 0.380}
\definecolor{y63}{rgb}{1.000, 1.000, 0.370}
\definecolor{y64}{rgb}{1.000, 1.000, 0.360}
\definecolor{y65}{rgb}{1.000, 1.000, 0.350}
\definecolor{y66}{rgb}{1.000, 1.000, 0.340}
\definecolor{y67}{rgb}{1.000, 1.000, 0.330}
\definecolor{y68}{rgb}{1.000, 1.000, 0.320}
\definecolor{y69}{rgb}{1.000, 1.000, 0.310}
\definecolor{y70}{rgb}{1.000, 1.000, 0.300}
\definecolor{y71}{rgb}{1.000, 1.000, 0.290}
\definecolor{y72}{rgb}{1.000, 1.000, 0.280}
\definecolor{y73}{rgb}{1.000, 1.000, 0.270}
\definecolor{y74}{rgb}{1.000, 1.000, 0.260}
\definecolor{y75}{rgb}{1.000, 1.000, 0.250}
\definecolor{y76}{rgb}{1.000, 1.000, 0.240}
\definecolor{y77}{rgb}{1.000, 1.000, 0.230}
\definecolor{y78}{rgb}{1.000, 1.000, 0.220}
\definecolor{y79}{rgb}{1.000, 1.000, 0.210}
\definecolor{y80}{rgb}{1.000, 1.000, 0.200}
\definecolor{y81}{rgb}{1.000, 1.000, 0.190}
\definecolor{y82}{rgb}{1.000, 1.000, 0.180}
\definecolor{y83}{rgb}{1.000, 1.000, 0.170}
\definecolor{y84}{rgb}{1.000, 1.000, 0.160}
\definecolor{y85}{rgb}{1.000, 1.000, 0.150}
\definecolor{y86}{rgb}{1.000, 1.000, 0.140}
\definecolor{y87}{rgb}{1.000, 1.000, 0.130}
\definecolor{y88}{rgb}{1.000, 1.000, 0.120}
\definecolor{y89}{rgb}{1.000, 1.000, 0.110}
\definecolor{y90}{rgb}{1.000, 1.000, 0.100}
\definecolor{y91}{rgb}{1.000, 1.000, 0.090}
\definecolor{y92}{rgb}{1.000, 1.000, 0.080}
\definecolor{y93}{rgb}{1.000, 1.000, 0.070}
\definecolor{y94}{rgb}{1.000, 1.000, 0.060}
\definecolor{y95}{rgb}{1.000, 1.000, 0.050}
\definecolor{y96}{rgb}{1.000, 1.000, 0.040}
\definecolor{y97}{rgb}{1.000, 1.000, 0.030}
\definecolor{y98}{rgb}{1.000, 1.000, 0.020}
\definecolor{y99}{rgb}{1.000, 1.000, 0.010}
\definecolor{y100}{rgb}{1.000, 1.000, 0.000}

\definecolor{w0}{rgb}{1.000, 1.000, 1.000}
\definecolor{w1}{rgb}{0.990, 0.990, 0.990}
\definecolor{w2}{rgb}{0.980, 0.980, 0.980}
\definecolor{w3}{rgb}{0.970, 0.970, 0.970}
\definecolor{w4}{rgb}{0.960, 0.960, 0.960}
\definecolor{w5}{rgb}{0.950, 0.950, 0.950}
\definecolor{w6}{rgb}{0.940, 0.940, 0.940}
\definecolor{w7}{rgb}{0.930, 0.930, 0.930}
\definecolor{w8}{rgb}{0.920, 0.920, 0.920}
\definecolor{w9}{rgb}{0.910, 0.910, 0.910}
\definecolor{w10}{rgb}{0.900, 0.900, 0.900}
\definecolor{w11}{rgb}{0.890, 0.890, 0.890}
\definecolor{w12}{rgb}{0.880, 0.880, 0.880}
\definecolor{w13}{rgb}{0.870, 0.870, 0.870}
\definecolor{w14}{rgb}{0.860, 0.860, 0.860}
\definecolor{w15}{rgb}{0.850, 0.850, 0.850}
\definecolor{w16}{rgb}{0.840, 0.840, 0.840}
\definecolor{w17}{rgb}{0.830, 0.830, 0.830}
\definecolor{w18}{rgb}{0.820, 0.820, 0.820}
\definecolor{w19}{rgb}{0.810, 0.810, 0.810}
\definecolor{w20}{rgb}{0.800, 0.800, 0.800}
\definecolor{w21}{rgb}{0.790, 0.790, 0.790}
\definecolor{w22}{rgb}{0.780, 0.780, 0.780}
\definecolor{w23}{rgb}{0.770, 0.770, 0.770}
\definecolor{w24}{rgb}{0.760, 0.760, 0.760}
\definecolor{w25}{rgb}{0.750, 0.750, 0.750}
\definecolor{w26}{rgb}{0.740, 0.740, 0.740}
\definecolor{w27}{rgb}{0.730, 0.730, 0.730}
\definecolor{w28}{rgb}{0.720, 0.720, 0.720}
\definecolor{w29}{rgb}{0.710, 0.710, 0.710}
\definecolor{w30}{rgb}{0.700, 0.700, 0.700}
\definecolor{w31}{rgb}{0.690, 0.690, 0.690}
\definecolor{w32}{rgb}{0.680, 0.680, 0.680}
\definecolor{w33}{rgb}{0.670, 0.670, 0.670}
\definecolor{w34}{rgb}{0.660, 0.660, 0.660}
\definecolor{w35}{rgb}{0.650, 0.650, 0.650}
\definecolor{w36}{rgb}{0.640, 0.640, 0.640}
\definecolor{w37}{rgb}{0.630, 0.630, 0.630}
\definecolor{w38}{rgb}{0.620, 0.620, 0.620}
\definecolor{w39}{rgb}{0.610, 0.610, 0.610}
\definecolor{w40}{rgb}{0.600, 0.600, 0.600}
\definecolor{w41}{rgb}{0.590, 0.590, 0.590}
\definecolor{w42}{rgb}{0.580, 0.580, 0.580}
\definecolor{w43}{rgb}{0.570, 0.570, 0.570}
\definecolor{w44}{rgb}{0.560, 0.560, 0.560}
\definecolor{w45}{rgb}{0.550, 0.550, 0.550}
\definecolor{w46}{rgb}{0.540, 0.540, 0.540}
\definecolor{w47}{rgb}{0.530, 0.530, 0.530}
\definecolor{w48}{rgb}{0.520, 0.520, 0.520}
\definecolor{w49}{rgb}{0.510, 0.510, 0.510}
\definecolor{w50}{rgb}{0.500, 0.500, 0.500}
\definecolor{w51}{rgb}{0.490, 0.490, 0.490}
\definecolor{w52}{rgb}{0.480, 0.480, 0.480}
\definecolor{w53}{rgb}{0.470, 0.470, 0.470}
\definecolor{w54}{rgb}{0.460, 0.460, 0.460}
\definecolor{w55}{rgb}{0.450, 0.450, 0.450}
\definecolor{w56}{rgb}{0.440, 0.440, 0.440}
\definecolor{w57}{rgb}{0.430, 0.430, 0.430}
\definecolor{w58}{rgb}{0.420, 0.420, 0.420}
\definecolor{w59}{rgb}{0.410, 0.410, 0.410}
\definecolor{w60}{rgb}{0.400, 0.400, 0.400}
\definecolor{w61}{rgb}{0.390, 0.390, 0.390}
\definecolor{w62}{rgb}{0.380, 0.380, 0.380}
\definecolor{w63}{rgb}{0.370, 0.370, 0.370}
\definecolor{w64}{rgb}{0.360, 0.360, 0.360}
\definecolor{w65}{rgb}{0.350, 0.350, 0.350}
\definecolor{w66}{rgb}{0.340, 0.340, 0.340}
\definecolor{w67}{rgb}{0.330, 0.330, 0.330}
\definecolor{w68}{rgb}{0.320, 0.320, 0.320}
\definecolor{w69}{rgb}{0.310, 0.310, 0.310}
\definecolor{w70}{rgb}{0.300, 0.300, 0.300}
\definecolor{w71}{rgb}{0.290, 0.290, 0.290}
\definecolor{w72}{rgb}{0.280, 0.280, 0.280}
\definecolor{w73}{rgb}{0.270, 0.270, 0.270}
\definecolor{w74}{rgb}{0.260, 0.260, 0.260}
\definecolor{w75}{rgb}{0.250, 0.250, 0.250}
\definecolor{w76}{rgb}{0.240, 0.240, 0.240}
\definecolor{w77}{rgb}{0.230, 0.230, 0.230}
\definecolor{w78}{rgb}{0.220, 0.220, 0.220}
\definecolor{w79}{rgb}{0.210, 0.210, 0.210}
\definecolor{w80}{rgb}{0.200, 0.200, 0.200}
\definecolor{w81}{rgb}{0.190, 0.190, 0.190}
\definecolor{w82}{rgb}{0.180, 0.180, 0.180}
\definecolor{w83}{rgb}{0.170, 0.170, 0.170}
\definecolor{w84}{rgb}{0.160, 0.160, 0.160}
\definecolor{w85}{rgb}{0.150, 0.150, 0.150}
\definecolor{w86}{rgb}{0.140, 0.140, 0.140}
\definecolor{w87}{rgb}{0.130, 0.130, 0.130}
\definecolor{w88}{rgb}{0.120, 0.120, 0.120}
\definecolor{w89}{rgb}{0.110, 0.110, 0.110}
\definecolor{w90}{rgb}{0.100, 0.100, 0.100}
\definecolor{w91}{rgb}{0.090, 0.090, 0.090}
\definecolor{w92}{rgb}{0.080, 0.080, 0.080}
\definecolor{w93}{rgb}{0.070, 0.070, 0.070}
\definecolor{w94}{rgb}{0.060, 0.060, 0.060}
\definecolor{w95}{rgb}{0.050, 0.050, 0.050}
\definecolor{w96}{rgb}{0.040, 0.040, 0.040}
\definecolor{w97}{rgb}{0.030, 0.030, 0.030}
\definecolor{w98}{rgb}{0.020, 0.020, 0.020}
\definecolor{w99}{rgb}{0.010, 0.010, 0.010}
\definecolor{w100}{rgb}{0.000, 0.000, 0.000}

\usepackage[natbib, bibencoding=utf8, citestyle=numeric, bibstyle=ieee, maxbibnames=999, maxcitenames=2, mincitenames=1, sortcites]{biblatex}
\bibliography{references}

\usepackage[acronym]{glossaries}
\newacronym{AC}{AC}{affective computing}
\newacronym{ANN}{ANN}{artificial neural network}
\newacronym{AI}{AI}{artificial intelligence}
\newacronym{ASR}{ASR}{automatic speech recognition}
\newacronym{CCC}{CCC}{concordance correlation coefficient}
\newacronym{CKA}{CKA}{centred kernel alignment}
\newacronym{CNN}{CNN}{convolutional neural network}
\newacronym{CRNN}{CRNN}{convolutional recurrent neural network}
\newacronym{CE}{CE}{cross entropy}
\newacronym{CI}{CI}{confidence interval}
\newacronym{COPD}{COPD}{chronic obstructive pulmonary disease}
\newacronym{CP}{CP}{computational paralinguistics}
\newacronym{CV}{CV}{cross-validation}
\newacronym{DANN}{DANN}{domain adversarial neural network}
\newacronym{DDPM}{DDPM}{denoising diffusion probabilistic model}
\newacronym{DL}{DL}{deep learning}
\newacronym{DNN}{DNN}{deep neural network}
\newacronym{DSP}{DSP}{digital signal processing}
\newacronym{eGeMAPS}{\emph{eGeMAPS}}{extended Geneva minimalistic acoustic descriptor set}
\newacronym{EU}{EU}{European Union}
\newacronym{EUAI}{EU AI Act}{European Union's Artificial Intelligence Act}
\newacronym{ERB}{ERB}{equivalent rectangular bandwidth}
\newacronym{ERM}{ERM}{empirical risk minimisation}
\newacronym{FFNN}{FFNN}{feed-forward neural network}
\newacronym{FFT}{FFT}{fast Fourier transform}
\newacronym{GAN}{GAN}{generative adversarial network}
\newacronym{GMM}{GMM}{Gaussian mixture model}
\newacronym{GRU}{GRU}{gated-recurrent unit}
\newacronym{HNR}{HNR}{harmonic-to-noise ratio}
\newacronym{HR}{HR}{heart rate}
\newacronym{IID}{IID}{independently and identically distributed}
\newacronym{IR}{IR}{impulse response}
\newacronym{ISWF}{ISWF}{isoelastic social welfare function}
\newacronym{LLD}{LLD}{low-level descriptor}
\newacronym{LLM}{LLM}{large language model}
\newacronym{LoRA}{LoRA}{low-rank adaptation}
\newacronym{LOSO}{LOSO}{leave-one-speaker-out}
\newacronym{LSTM}{LSTM}{long short-term memory network}
\newacronym{MAC}{MAC}{multiply-addition computation}
\newacronym{MIR}{MIR}{microphone impulse response}
\newacronym{ML}{ML}{machine learning}
\newacronym{MLP}{MLP}{multilayered perceptron}
\newacronym{MSE}{MSE}{mean squared error}
\newacronym{NLP}{NLP}{natural language processing}
\newacronym{NN}{NN}{neural network}
\newacronym{NLLoss}{NLLoss}{nonnegative likelihood loss}
\newacronym{OOD}{OOD}{out-of-domain}
\newacronym{PANN}{PANN}{pretrained audio neural network}
\newacronym{RBF}{RBF}{radial basis function}
\newacronym{ResNet}{ResNet}{residual neural network}
\newacronym{RIR}{RIR}{room impulse response}
\newacronym{RMSE}{RMSE}{root mean squared error}
\newacronym{RNN}{RNN}{recurrent neural network}
\newacronym{SDR}{SDR}{signal-to-distortion ratio}
\newacronym{SED}{SED}{sound event detection}
\newacronym{SER}{SER}{speech emotion recognition}
\newacronym{SGD}{SGD}{stochastic gradient descent}
\newacronym{SI-SDR}{SI-SDR}{scale-invariant signal-to-distortion ratio}
\newacronym{SNR}{SNR}{signal-to-noise ratio}
\newacronym{SSL}{SSL}{self-supervised learning}
\newacronym{STFT}{STFT}{short-time Fourier transform}
\newacronym{SVM}{SVM}{support vector machine}
\newacronym{TDNN}{TDNN}{time-delay neural network}
\newacronym{TPR}{TPR}{true positive rate}
\newacronym{TTS}{TTS}{text-to-speech}
\newacronym{UAR}{UAR}{unweighted average recall}
\newacronym{ZSL}{ZSL}{zero-shot learning}

\newcommand\cnn{\mbox{\emph{CNN14}}}
\newcommand\effnet{\emph{EfficientNet}}
\newcommand\wbase{\mbox{\emph{w2v2-b}}}
\newcommand\hbase{\mbox{\emph{hubert-b}}}
\newcommand\wlarge{\mbox{\emph{w2v2-L}}}
\newcommand\hlarge{\mbox{\emph{hubert-L}}}
\newcommand\wrobust{\mbox{\emph{w2v2-L-robust}}}

\newcommand\wvox{\mbox{\emph{w2v2-L-vox}}}

\newcommand\wemo{\mbox{\emph{w2v2-L-12-avd}}}
\newcommand\wser{\mbox{\emph{w2v2-L-12-emo}}}

\newcommand{\wav}{\emph{w2v2}}

\newcommand{\bert}{\emph{BERT}}
\newcommand{\hubert}{\emph{hubert}}
\newcommand{\opensmile}{\emph{openSMILE}}

\newcommand{\aibo}{\emph{FAU-AIBO}}
\newcommand{\dcase}{\emph{TAU-UAVS}}
\newcommand{\iemocap}{\emph{IEMOCAP}}
\newcommand{\podcast}{\emph{MSP-Podcast}}
\newcommand{\podcastseven}{\emph{MSP-Podcast-v1.7}}

\newcommand{\podcasteleven}{\emph{MSP-Podcast-v1.11}}
\newcommand{\emodb}{\emph{EmoDB}}

\newcommand{\ie}{i.\,e.\,,}

\usepackage[capitalize]{cleveref}
\begin{document}
  
\title{Charting 15 years of progress in deep learning for speech emotion recognition: A replication study

\thanks{\textit{\underline{Citation}}: 
\textbf{Authors. Title. Pages.... DOI:000000/11111.}} 
}

\author{
  \textbf{Andreas Triantafyllopoulos}$^{1,2}$, \textbf{Anton Batliner}$^{1,2}$, \textbf{Björn W. Schuller}$^{1,2,3,4}$\\
  $^1$CHI -- Chair of Health Informatics, Technical University of Munich, Munich, Germany\\
  $^2$MCML -- Munich Center for Machine Learning, Munich, Germany\\
  $^3$MDSI -- Munich Data Science Institute, Munich, Germany\\
  $^4$GLAM -- Group on Language, Audio, \& Music, Imperial College, London, UK\\
  \texttt{andreas.triantafyllopoulos@tum.de}
}

\maketitle
\begin{abstract}
Speech emotion recognition (SER) has long benefited from the adoption of deep learning methodologies.
Deeper models -- with more layers and more trainable parameters -- are generally perceived as being `better' by the SER community.
This raises the question -- \emph{how much better} are modern-era deep neural networks compared to their earlier iterations?
Beyond that, the more important question of how to move forward remains as poignant as ever.
SER is far from a solved problem; therefore, identifying the most prominent avenues of future research is of paramount importance.
In the present contribution, we attempt a quantification of progress in the 15 years of research beginning with the introduction of the landmark 2009 INTERSPEECH Emotion Challenge.
We conduct a large scale investigation of model architectures, spanning both audio-based models that rely on speech inputs and text-baed models that rely solely on transcriptions.
Our results point towards diminishing returns and a plateau after the recent introduction of transformer architectures.
Moreover, we demonstrate how perceptions of progress are conditioned on the particular selection of models that are compared.
Our findings have important repercussions about the state-of-the-art in SER research and the paths forward\footnote{Code repository: \url{https://github.com/CHI-TUM/ser-progress-replication}}.
\end{abstract}


\section{Introduction}

\IEEEPARstart{D}{eep} learning has arguably become the `go-to' method for \gls{SER} in the past decade.
Featured prominently in recent publications and surveys~\citep{Schuller18-SER, Khalil19-SER}, the overwhelming consensus in the community is that \gls{DL} outperforms previous methods.
For instance, all submissions to the recent \emph{Odyssey 2024} \gls{SER} challenge relied heavily on \gls{DL} models to either extract learnt representations or fine-tune them on the challenge dataset~\citep{Goncalves24-O2S}.

\Glspl{ANN} have been trained for \gls{SER} since as early as 1998 with \citet{Huber98-YBM} (\gls{MLP} for two classes), see as well \citet{Nicholson00-ERI} who used a shallow, 2-layer \gls{MLP} in 2000.
The field has come a long way since then, with one key trend being the transition from handcrafted features towards end-to-end pipelines that incorporate feature extraction and modelling within a single, monolithic architecture~\citep{Trigeorgis16-AFE}, and another being the progression towards \emph{deeper} and \emph{larger} methods ({\ie} with more layers and more parameters)~\citep{Schuller18-SER, Wagner23-DOT}.
Some representative examples of this trend are the early work of \citet{Trigeorgis16-AFE}, who relied on \glspl{CRNN} operating on raw audio, and \citet{Wagner23-DOT} who demonstrated the superiority of transformer models that accept as input raw audio and have been pre-trained using \gls{SSL}.

However, with dozens -- if not hundreds -- of papers published yearly on the topics of \gls{SER} and \gls{DL}, it is hard to judge which advances (if any) substantially progress beyond the state-of-the-art -- and why.
Recent benchmark studies, such as \emph{HEAR}~\citep{Turian22-HHE} and \emph{SUPERB}~\citep{Yang21-SSP}, focus only tangentially on \gls{SER}, with both featuring single, small, and acted \gls{SER} datasets (\emph{CREMA-D}~\citep{Cao14-CCE} and \emph{IEMOCAP}~\citep{Busso08-IIE}, respectively), and only benchmarking recent \gls{DL} methods.
While their contribution to the general speech and audio community is undisputed, they fail to capture the progress that has been made on \gls{SER} for \emph{naturalistic} conditions.
Moreover, as these studies are focused on benchmarking, they are limited with respect to their analysis of why some models perform better than others.
This prevents them from uncovering important insights that can help guide future advances.
For example, \citet{Wagner23-DOT} showed that transformer models provide substantial improvements to valence recognition, but this was later attributed to their implicit modelling of linguistic content, rather than the discovery of new, unknown paralinguistic concepts related to valence~\citep{Triantafyllopoulos22-PSE}.
On the other hand, recurring challenge series, such as \emph{ComParE}~\footnote{http://www.compare.openaudio.eu/} and \emph{Odyssey}, are more suited to capturing the \emph{Zeitgeist} at specific moments in time, as participants compete with one another to obtain the best results using the latest methods, without necessarily revisiting older ones.

The problem is exacerbated by the fact that the field of \gls{SER} is lacking a universally-accepted benchmark of speech collected in naturalistic 
conditions.
{\aibo}, the dataset adopted by the 2009 INTERSPEECH Emotion Challenge -- the world's first \gls{SER} challenge -- was quietly abandoned over the years with very few papers using it in its original form~\citep{Triantafyllopoulos24-I2E}.
Similarly, the largest naturalistic \gls{SER} dataset, {\podcast}~\citep{Lotfian19-BNE}, has seen several releases in the last few years, making comparisons across different works harder.
Instead, the most widely-cited datasets became {\iemocap}~\citep{Busso08-IIE} and {\emodb}~\citep{Burkhardt05-ADO}.  {\iemocap} features only acted speech (both scripted and improvised) and lacks an established evaluation scheme (with some works relying on leave-one-speaker-out cross-validation and others preferring leave-one-session-out). {\emodb}, which is also acted, lacks a common evaluation scheme, and features only 
restricted
linguistic content. 
(Note that {\emodb} was originally intended to lay the foundations for emotional speech synthesis, not to  train  \gls{SER}.)  

As such, \gls{SER} is suffering from a `blind spot' as pertains to the progress that has been made by switching to larger and deeper models over the years.
This is a blind spot we aimed to address in our recent comparison of different models on the 2009 INTERSPEECH Emotion Challenge data~\citep{Triantafyllopoulos24-I2E}.
In the present work, we extend this study by additionally including {\podcast} as another, more recent and naturalistic, dataset of emotional speech.
Furthermore, we present results for \glspl{LLM} using textual data, and supplement our performance comparison with several analyses to better understand the progress made by using \gls{DL} architectures.

The remainder of our work is organised as follows.
In \cref{sec:related}, we present relevant background, discussing benchmarking approaches and offering an overview of the most important \gls{DL} trends seen in \gls{SER} research in the last decade.
\cref{sec:methodology} outlines our methodology and \cref{sec:results} showcases our results.
\cref{sec:discussion} includes a discussion of our findings and our conclusion, \cref{sec:limitations} the limitations of our work, and \cref{sec:future} our suggestions for future work.

\section{Related work}
\label{sec:related}
This section presents related work, beginning with a discussion of benchmarking in the context of \gls{SER} and continuing with a succinct overview of the main \gls{DL} architecture families employed for \gls{SER}.

\subsection{Benchmarking}

As discussed in \cref{sec:background:dls}, there is a variety of architectures that have been introduced for speech analysis, and particularly for \gls{SER}, but these are often evaluated on only a small number of datasets, with no consensus on common benchmarks.
Typically, a single work will refer to only a few `baseline' methods from prior work, oftentimes without reproducing the results but taking them from the literature.
This makes it hard to compare different models and algorithms proposed by different groups.
Crucially, it makes it hard to quantify the progress that has been made as new classes of architectures, like \glspl{CNN} or transformers, are introduced.

The problem is further exacerbated by a gradual change in the datasets used by the majority of the community (which often, but not always, occurs for good reasons).
For example, the first large-scale dataset open-sourced for \gls{SER} research was \aibo~\citep{Batliner08-RAT}, which was used for the 2009 INTERSPEECH Emotion Challenge~\citep{Schuller09-TI2a}.
While large (for its time) and well-known in the community, it was largely abandoned in the latter half of the 2010s with very few works using it (and even less in the form published for the challenge).
Instead, the community transitioned to other datasets, such as \iemocap~\citep{Busso08-IIE} (the top-cited \gls{SER} dataset at the time of writing), \emph{SEWA}~\citep{Kossaifi19-SDA}, \podcast~\citep{Lotfian19-BNE}, or others.
We note that this change was not always motivated by a quest for more `naturalness' or `bigger' data.
Indeed, {\aibo} is both larger and collected in a more realistic, not explicitly prompted setting than \iemocap~\citep{Busso08-IIE}, \emph{RAVDESS}~\citep{Livingstone18-TRA}, \emph{CREMA-D}~\citep{Cao14-CCE} and other datasets which have become more prevalent in recent years.

Moreover, the matter of properly benchmarking \gls{AI} models is far from trivial.
Beyond the choice of data discussed above, it is important to define an appropriate \emph{computational budget} and overall \emph{engineering effort} devoted to each particular method.
For example, the winner of the 2009 INTERSPEECH Emotion Challenge utilised a hierarchical task tree by identifying supergroups of target emotions using expert knowledge~\citep{Lee11-ERU}.
While this is a valid -- and very innovative -- approach, it is definitely more byzantine than the simple baseline which relied on `mere' classification.
Similar `over-engineered' approaches have been proposed over the years; from ones that take into account the exact problem proposed by a dataset, to ones with elaborate feature engineering.
These require substantial effort to reproduce, making their inclusion in a benchmarking effort prohibitively time-consuming.
In our present work, we chose to abide by the ``bitter lesson'' of \citet{Sutton19-TBL}, namely, that simpler approaches will scale better given more computational resources.
This fits the broader landscape of contemporary \gls{SER} research -- that of almost complete dominance of \gls{DL}.
Therefore, we focused on `simple' \gls{DNN} models and trained each of them with a fixed computational budget (in the form of update steps, not compute time) and hyperparameters.
While not exhaustive, and potentially disadvantaging some approaches, this allows for a more fair comparison.

\subsection{SER model timeline}
\label{sec:background:dls}
\gls{DNN} architectures
used for \gls{SER}
evolved substantially throughout the years.
In the following,
we present a list
of the most popular \gls{SER} models of recent years,
beginning with the launch
of the 2009 INTERSPEECH Emotion Challenge~\citep{Schuller09-TI2a}
and continuing to present day.
As the most prominent task in \gls{CP} research,
innovation in \gls{SER} has largely driven other related subfields as well,
and is thus representative of the field as a whole.

\noindent
\textbf{'09-: {\opensmile} feature sets} -- In the different iterations of the ComParE Challenge, newer versions of paralinguistic features were introduced.
In general, these feature sets were larger and covered a wider gamut of acoustic and prosodic features than the initial features introduced in the 2009 INTERSPEECH Emotion Challenge.
However, that period saw a parallel pursuit for \emph{smaller} expert-driven feature sets~\citep{Eyben15-TGM}.
Researchers used either the \emph{dynamic} low-level descriptors or their higher-order \emph{static} functionals.
These features were fed into \glspl{MLP} or \glspl{RNN} for further processing.

\noindent
\textbf{'12-: \emph{ImageNet} pre-training} -- Following the introduction of \emph{ImageNet} and the first \glspl{CNN} trained on it in 2012, such networks were subsequently introduced in the audio and speech domains by substituting images with (pictorial representations of) spectrograms~\citep{Cummins17-AID, Zhao19-EDS} -- a practice that is relevant to this day, with audio transformer models oftentimes initialised with states pre-trained on \emph{ImageNet}~\citep{Gong21-AAS}.

\noindent
\textbf{'16-: End-to-end} -- Subsequent years saw the introduction of \emph{end-to-end} models, \ie{} models trained directly on raw audio input for the target task without any prior feature pre-processing~\citep{Trigeorgis16-AFE}.
These models were especially successful in the case of time-continuous \gls{SER}, which requires predicting the emotion of very short audio frames, and usually follow the \gls{CRNN} architecture.

\noindent
\textbf{'16-: Supervised audio pre-training} -- In parallel to \emph{ImageNet} pre-training, there were also efforts to collect similar large-scale datasets for audio where networks could be pre-trained in a supervised fashion.
Two notable examples are \emph{VoxCeleb} and \emph{AudioSet}~\citep{Gemmeke17-ASA}, both collected from YouTube, with the former targeted to speaker identification and the latter to general audio tagging.
\emph{VoxCeleb} formed the basis for training speaker embedding models (\ie{} `x-vectors') using \glspl{TDNN}~\citep{Desplanques20-EEC}.
\emph{AudioSet} in turn inspired the use of convolutional networks, such as \emph{CNN14}, which was introduced in \glspl{PANN}~\citep{Kong20-PLS} and shown to also transfer well to \gls{SER} tasks~\citep{Triantafyllopoulos21-TRO}, and later, transformer-based models such as \emph{AST}~\citep{Gong21-AAS}.
In addition, the introduction of the \emph{Whisper} architecture~\citep{Radford23-RSR} 
led to a renaissance of supervised training for \gls{ASR}.

\noindent
\textbf{'20-: Self-supervised audio pre-training} -- The introduction of transformers and the advent of self-supervised pre-training for computer vision and \gls{NLP} also propagated to the speech and audio domain.
The two dominant architectures here are \emph{wa2v2ec2.0}(\wav)~\citep{Baevski20-W2V} and \hubert~\citep{Hsu21-HSS}.
These models have yielded significant advances in \gls{SER}, as also shown in our own work~\citep{Wagner23-DOT, Triantafyllopoulos24-I2E}, which we have partially accredited to their ability to simultaneously encode linguistic and paralinguistic information~\citep{Triantafyllopoulos22-PSE}.

\section{Methodology}
\label{sec:methodology}
This section outlines our workflow, beginning with the considered datasets, continuing with the experimental settings used for training models, and finishing with the details of the accompanying analyses.

\subsection{Datasets and tasks}

As mentioned, we employed two different datasets.
See \cref{app:data} for a broader discussion on our dataset selection criteria.
{\aibo} is the official dataset of the INTERSPEECH 2009 Emotion Challenge~\citep{Schuller09-TI2a}.
It is a dataset of German children's speech collected in a Wizard-of-Oz scenario and annotated on the word-level for the presence of $11$ emotional/communicative states by $5$ expert raters~\citep{Batliner04-YST, Steidl09-ACO}.
Subsequently, segmented words have been aggregated to meaningful chunks using manual semantic and prosodic criteria.
Accordingly, annotated states have been mapped to $2$- and $5$-class categorisations with a set of heuristics, which forms a final dataset of $18\,216$ chunks used for the challenge.
\aibo$^2$ features the classes \emph{negative} (NEG) and \emph{non-negative} (IDL), whereas \aibo$^5$ comprises \emph{anger} (A), \emph{emphatic} (E), \emph{neutral} (N), \emph{motherese} (P), and \emph{rest} (R) for the remaining classes.
The data is heavily imbalanced towards the neutral/non-negative classes.
The data was collected from two schools, with one school set aside for testing (\emph{Mont}) and one set aside for training (\emph{Ohm}); we use the same partitioning for our experiments.
Additionally, we create a small validation set comprising the last two speakers of the training set (speakers are denoted by number IDs): \emph{Ohm\_31} and \emph{Ohm\_32} as in our previous work~\citep{Triantafyllopoulos24-I2E}.
Note that the data are extensively documented in \citet{Steidl09-ACO}.

{\podcast}~\citep{Lotfian19-BNE} is a recently-introduced data set for \gls{SER}, and one of the biggest publicly-released datasets to date.
However, one `downside' of it is that new releases are made approximately every year; these releases include increases to the data size but also corrections to the annotations and files included.
We used {\podcasteleven} as the latest version of the dataset available to us, comprising $44\,586$ instances in the training set, $11\,947$ in the validation, and $20\,845$ in the test set.
The dataset has been annotated for the emotional dimensions of \emph{arousal}, \emph{valence}, \emph{dominance}, as well as for the emotion categories of \emph{angry}, \emph{contemptful}, \emph{disgusted}, \emph{afraid}, \emph{happy}, \emph{neutral}, \emph{sad}, and \emph{surprised}.
A 7-point Likert scale was used for the emotional dimensions on the utterance level,
and scores by individual annotators have been averaged to obtain a consensus vote.
All versions are split into speaker independent partitions with an official train/dev/test partition; some releases feature a \emph{test-2} partition which we ignore throughout our work.

\subsection{Experimental settings}

In total, we experimented with $43$ audio-based models (see \cref{app:models} for the complete list).
All audio models were trained using \emph{autrainer}~\citep{Rampp24-AAM}.
To constrain our space of hyperparameters, we conducted two experimental phases.
In the first \textbf{exploration phase}, we tested all $43$ investigated models using a fixed set of hyperparameters.
Specifically, we used the \emph{Adam} optimiser with a learning rate of $.0001$ for $30$ epochs.
We used a batch size of $4$ due to hardware constraints (access to higher-end GPUs that could fit all models at higher batch sizes was more limited to us than lower-end GPUs which fit all models at a batch size of $4$).
In total, this resulted in $43$ runs for the exploration phase of each task.

Additionally, we conducted a \textbf{tuning phase}, where we further optimised a larger set of hyperparameters for the $5$ best-performing models from the exploration phase for \aibo$^2$, \aibo$^5$, and {\podcasteleven}, doing a grid search over optimisers \{\emph{Adam}, \emph{SGD}\} and learning rates $\{.01,.001,.0001\}$, while training each configuration for $50$ epochs for {\aibo} and $5$ epochs for {\podcasteleven} (due to time constraints).
For {\aibo}, we additionally explored different batch sizes $\{4,8,16\}$.

To account for variable-length sequences in training, we randomly cropped/padded all chunks to a fixed length of $3$ seconds (different cropping/padding was applied on each instance across different epochs; random seed was fixed and can be reproduced) when using dynamic features (including the raw audio); during inference, we used the original utterances, only padding those shorter than $2$ seconds with silence.
Except for {\wav} and {\hubert}, where we froze the feature extractors, all model parameters were fine-tuned.
We used the weighted cross-entropy loss for classification to account for class imbalance.
In all cases, we used the defined validation set to select the best-performing epoch for each model, which we then evaluated on the test set.

Our search over text-based models was much more restricted.
It was also inspired by recent developments in \gls{NLP},
but was much more constrained in time.
We evaluated {\bert}~\citep{Devlin18-BPO},
\emph{RoBERTa}~\citep{Liu19-RAR},
\emph{DistilBERT}~\citep{Sanh19-DAD},
and \emph{Electra}~\citep{Clark20-EPT}
as ``first generation'' transformers,
and \emph{Llama-2}~\citep{Touvron23-L2O}, \emph{Llama-3}~\citep{Grattafiori24-TL3}, and \emph{Mistral}~\citep{Jiang24-M7B}
as ``second generation'' \glspl{LLM}.
First generation transformers
were trained with a batch size of $8$,
\emph{AdamW} with a weight decay of $.00001$,
and a learning rate of $.00001$
for $10$ epochs.
Second generation \glspl{LLM}
were trained for $5$ epochs
with a learning rate of $.00001$,
\emph{AdamW} with a weight decay of $.01$,
and \gls{LoRA} fine-tuning
of only the attention projection matrices
(for query, key, value, and output)~\citep{Hu22-LLA}
with $4$-bit quantisation of model weights.
For {\aibo},
we used the official transliteration
provided by human raters.
For {\podcasteleven},
we used automatic transcriptions
generated with \emph{Whisper-large-v2}.

\subsection{Model evaluation}

\textbf{\Gls{IID} evaluation:} Models were initially evaluated on the corresponding test set of each dataset.
We report \gls{UAR} -- the standard metric for \gls{SER} since the 2009 INTERSPEECH Emotion Challenge which accounts for class imbalance by computing first the recall per class and then averaging.

\textbf{\Gls{OOD} evaluation:} Similar to \citet{Wagner23-DOT}, we also evaluated \gls{OOD} performance.
To do so, we used the \gls{SER} models trained on {\podcasteleven}, as this is the only task for which it was possible to find \gls{OOD} data (due to the non-standard labels used in {\aibo}).
Specifically, we used {\emodb}~\citep{Burkhardt05-ADO} and {\iemocap}~\citep{Busso08-IIE}.
For both datasets, we used the entire data corresponding to the four classes that the model was trained on.
For {\iemocap}, we additionally re-labelled samples belonging to the ``excited'' class as ``happy''.
Performance was once again evaluated using \gls{UAR}.

\subsection{Probing features}
In order to better understand 
which paralinguistic features
models rely on
-- especially after fine-tuning --
we searched for interpretable features
that are predictable
from the intermediate representations
of transformer models.
This is known as \emph{probing}~\citep{Triantafyllopoulos22-PSE}.
For our probing experiments,
we relied on a small, easily-interpretable subset
of the \gls{eGeMAPS}~\citep{Eyben15-TGM}.
Those were:
\underline{$\mu$(P)} average pitch; \underline{$\sigma$(P)} standard deviation of pitch; \underline{$\mu$(L)} average loudness; \underline{$\mu$(L)\,[dB]} average loudness in dB; \underline{$\mu$(J)} average jitter; \underline{$\mu$(S)} average shimmer; \underline{$\mu$(HNR)} average harmonic-to-noise ratio; \underline{$\mu$(F1)} average frequency for first formant; \underline{$\mu$(F2)} average frequency for second formant; \underline{$\mu$(F3)} average frequency for third formant.

Concretely,
we extracted hidden representations
from all layers
and all transformer models 
that were fine-tuned on {\podcasteleven},
namely,
{\wbase}, {\wlarge}, {\wvox}, {\wrobust}, {\wemo},
{\hbase}, and {\hlarge}.
We then averaged these representations over the time dimension.
Subsequently,
we used a linear regression model
to predict the absolute value of each feature
scaled in the range $[0, 1]$.
Models were trained and evaluated
using \gls{LOSO} \gls{CV} on {\emodb}
and we measured their Pearson's $r$.

\subsection{Robustness analysis}
For our investigation of robustness,
we created a noisy mixture of {\podcasteleven},
leveraging the 2021 version 
of the TAU Urban Audio-Visual Scenes Development dataset 
(\dcase)~\citep{Wang21-ACD}
as our source for additive noise.
All audio was sourced 
from its \emph{training set}, 
where we used the first $10$ instances 
of each label in each city 
to create our mixtures.
Specifically,
for every test instance of {\podcasteleven},
we randomly drew one audio file
from {\dcase}
with replacement,
which we added to the original audio
from {\podcasteleven}
using a \gls{SNR} of $0$\,dB.
We then evaluated
models trained on clean audio
to measure the deterioration in performance
caused by additive noise.

\subsection{Individual fairness}
The standard process for computing performance in \gls{ML} tasks is to consider each instance as an \emph{independent trial}.
Model performance is then measured using some metric, $\mathcal{M}$, over the set of reference labels $\{y\}_0^N$ 
and predictions $\{\hat{y}\}_0^N$, with $N$ being the size of the evaluation set.
However, the disparity in performance across different individuals plays a vital role in \gls{SER}~\citep{Triantafyllopoulos24-EPF}.
Therefore, it makes sense to consider \emph{individual-level performance}, and, as a corollary, \emph{individual-level fairness}.

To define individual-level fairness, we began by computing the performance on a speaker-level, thus assuming that speakers are first selected independently, and only then are samples selected independently for them (see \citet{Guyon98-WST} for a similar argumentation).
We finally computed the metric $\mathcal{M}$ over the set of instances for each individual speaker in our dataset, which we denote as $SP_{\mathcal{M}}$ in order to differentiate it from the standard evaluation protocol of treating each utterance independently, which we denote by $U_{\mathcal{M}}$.
We call $SP_{\mathcal{M}}$ the \textbf{utility} of each speaker, as this is the benefit that each speaker can expect from getting their emotions recognised correctly\footnote{Naturally, what this `benefit' entails depends entirely on the downstream application.}.
We examined this utility by taking the Gini coefficient ($G(\cdot)$) as a standard measurement used in econometrics to judge the distribution of utility in a particular population; it is defined as half of the mean absolute difference relative to the mean of a particular sample~\citep{Dorfman79-AFF}.
The Gini coefficient has been previously used to quantify inequality in speaker-level performance for \gls{SER}~\citep{Triantafyllopoulos24-EPF}.
In particular, it takes the value of $0$ for an equal distribution where everyone has the same utility, and $1$ for a completely unequal one, where the entire utility is accumulated by one particular individual.
In our case, this was computed as:
\begin{equation}
    G(\{u_i\}_1^N) = \frac{\sum\limits_{i=1}^{i=N}\left(\sum\limits_{j=1}^{j=N}(|u_i-u_j|)\right)}{\mu(\{u_i\}_1^N)},
\end{equation}
where $N$ is the number of speakers in the test set and $\{u_i\}_1^N$ is the set of utility values for all speakers, with $u_i = SP_{\mathcal{M}}^i$, {\ie} the speaker-level performance computed for each speaker.

\section{Results}
\label{sec:results}
This section presents our performance comparisons for all audio models both \gls{IID} and \gls{OOD}.
Moreover, it contains our probing analysis and results on robustness and individual fairness.

\subsection{Audio-based models}
\label{sec:audio}
We begin the presentation of our results by exclusively considering audio-based models, {\ie} models that accept speech (either in raw form or as features).
A complete list of models and pre-trained states is provided in \cref{app:models}.

\subsubsection{IID results}

\begin{table}
    \centering
    \caption{
    \Gls{UAR} results for all speech-based models in the \textbf{exploration phase} using the hyperparameters defined above (\emph{Adam}, $0.0001$, $4$/$16$) for both the $2$- and the $5$-class tasks of the 2009 INTERSPEECH Emotion Challenge using the {\aibo} dataset, as well as the $4$-class categorical \gls{SER} task of \podcasteleven.
    }
    \label{tab:exploration}
   \begin{tabular}{lrrr}
    \toprule
    & \aibo$^2$ & \aibo$^5$ & \podcasteleven \\
    \midrule
    \emph{IS09-s-mlp} & \cellcolor{w23}{.620} & \cellcolor{w1}{.294} & \cellcolor{w0}{.250} \\
    \emph{IS09-d-lstm} & \cellcolor{w48}{.670} & \cellcolor{w25}{.347} & \cellcolor{w21}{.469} \\
    \emph{IS10-s-mlp} & \cellcolor{w49}{.670} & \cellcolor{w25}{.348} & \cellcolor{w31}{.504} \\
    \emph{IS10-d-lstm} & \cellcolor{w58}{.689} & \cellcolor{w50}{.406} & \cellcolor{w26}{.486} \\
    \emph{IS11-s-mlp} & \cellcolor{w0}{.499} & \cellcolor{w0}{.201} & \cellcolor{w30}{.502} \\
    \emph{IS11-d-lstm} & \cellcolor{w48}{.670} & \cellcolor{w36}{.373} & \cellcolor{w29}{.499} \\
    \emph{IS12-s-mlp} & \cellcolor{w0}{.501} & \cellcolor{w0}{.201} & \cellcolor{w30}{.501} \\
    \emph{AlexNet} & \cellcolor{w0}{.503} & \cellcolor{w0}{.200} & \cellcolor{w0}{.250} \\
    \emph{IS12-d-lstm} & \cellcolor{w53}{.679} & \cellcolor{w36}{.373} & \cellcolor{w28}{.496} \\
    \emph{IS13-s-mlp} & \cellcolor{w0}{.498} & \cellcolor{w0}{.201} & \cellcolor{w30}{.503} \\
    \emph{IS13-d-lstm} & \cellcolor{w45}{.664} & \cellcolor{w38}{.378} & \cellcolor{w31}{.505} \\
    \emph{eGeMAPS-s-mlp} & \cellcolor{w22}{.618} & \cellcolor{w0}{.245} & \cellcolor{w24}{.479} \\
    \emph{eGeMAPS-d-lstm} & \cellcolor{w47}{.667} & \cellcolor{w46}{.397} & \cellcolor{w27}{.491} \\
    \emph{Resnet50} & \cellcolor{w58}{.688} & \cellcolor{w60}{.428} & \cellcolor{w32}{.510} \\
    \emph{VGG}$^{19}$ & \cellcolor{w0}{.499} & \cellcolor{w0}{.204} & \cellcolor{w0}{.250} \\
    \emph{IS16-s-mlp} & \cellcolor{w0}{.500} & \cellcolor{w0}{.201} & \cellcolor{w29}{.497} \\
    \emph{VGG}$^{16}$ & \cellcolor{w36}{.646} & \cellcolor{w49}{.404} & \cellcolor{w22}{.471} \\
    \emph{IS16-d-lstm} & \cellcolor{w41}{.655} & \cellcolor{w40}{.382} & \cellcolor{w29}{.498} \\
    \emph{VGG}$^{13}$ & \cellcolor{w46}{.665} & \cellcolor{w41}{.385} & \cellcolor{w0}{.250} \\
    \emph{VGG}$^{11}$ & \cellcolor{w46}{.666} & \cellcolor{w0}{.200} & \cellcolor{w18}{.459} \\
    \emph{CRNN}$^{18}$ & \cellcolor{w53}{.680} & \cellcolor{w44}{.392} & \cellcolor{w33}{.514} \\
    {\effnet} & \cellcolor{w48}{.669} & \cellcolor{w27}{.353} & \cellcolor{w40}{.537} \\
    \emph{CRNN}$^{19}$ & \cellcolor{w55}{.683} & \cellcolor{w35}{.372} & \cellcolor{w30}{.503} \\
    \wlarge & \cellcolor{w0}{.500} & \cellcolor{w0}{.200} & \cellcolor{w0}{.250} \\
    \wbase & \cellcolor{w0}{.500} & \cellcolor{w0}{.200} & \cellcolor{w0}{.250} \\
    \emph{ConvNeXt}$^t$ & \cellcolor{w0}{.500} & \cellcolor{w48}{.400} & \cellcolor{w35}{.518} \\
    \emph{ConvNeXt}$^l$ & \cellcolor{w45}{.663} & \cellcolor{w52}{.409} & \cellcolor{w35}{.518} \\
    \emph{ConvNeXt}$^b$ & \cellcolor{w46}{.665} & \cellcolor{w53}{.412} & \cellcolor{w33}{.513} \\
    \emph{ConvNeXt}$^s$ & \cellcolor{w51}{.674} & \cellcolor{w50}{.406} & \cellcolor{w38}{.531} \\
    \emph{ETDNN} & \cellcolor{w53}{.678} & \cellcolor{w50}{.404} & \cellcolor{w37}{.528} \\
    \cnn & \cellcolor{w60}{.692} & \cellcolor{w45}{.394} & \cellcolor{w36}{.524} \\
    \hbase & \cellcolor{w0}{.500} & \cellcolor{w0}{.200} & \cellcolor{w0}{.250} \\
    \emph{Swin}$^b$ & \cellcolor{w0}{.528} & \cellcolor{w0}{.200} & \cellcolor{w0}{.250} \\
    \emph{Swin}$^t$ & \cellcolor{w0}{.530} & \cellcolor{w0}{.242} & \cellcolor{w0}{.269} \\
    \emph{AST} & \cellcolor{w0}{.535} & \cellcolor{w4}{.300} & \cellcolor{w1}{.396} \\
    \wvox & \cellcolor{w33}{.640} & \cellcolor{w48}{.402} & \cellcolor{w47}{.561} \\
    \hlarge & \cellcolor{w47}{.667} & \cellcolor{w56}{.418} & \cellcolor{w49}{.570} \\
    \emph{Swin}$^s$ & \cellcolor{w50}{.672} & \cellcolor{w7}{.306} & \cellcolor{w0}{.250} \\
    \wrobust & \cellcolor{w56}{.684} & \cellcolor{w52}{.411} & \cellcolor{w45}{.555} \\
    \emph{Whisper}$^s$ & \cellcolor{w41}{.656} & \cellcolor{w0}{.279} & \cellcolor{w3}{.405} \\
    \wemo & \cellcolor{w55}{.684} & \cellcolor{w53}{.411} & \cellcolor{w60}{.609} \\
    \emph{Whisper}$^t$ & \cellcolor{w56}{.684} & \cellcolor{w39}{.380} & \cellcolor{w0}{.387} \\
    \emph{Whisper}$^b$ & \cellcolor{w57}{.686} & \cellcolor{w0}{.200} & \cellcolor{w0}{.250} \\
    \bottomrule
    \end{tabular}
\end{table}

Results for the \emph{exploration phase} are shown in \cref{tab:exploration}.
For {\aibo}, the best-performing model for the $2$-class problem was \cnn, with a \gls{UAR} of $.692$, whereas for the $5$-class problem it was \emph{ResNet50}, with a \gls{UAR} of $.428$.
We further observe that some models failed to converge and yield chance (or near-chance) performance -- most likely caused by the choice of hyperparameters, which favour some models more than others.
Notably, most of our results were below the challenge winners for {\aibo} (\gls{UAR}: $.703/.417$) and several were in the same `ballpark' as the original baseline (\gls{UAR}: $.677/.382$).
For \podcasteleven, the best performance was achieved by {\wemo} with a \gls{UAR} of $.609$; however, this model has been already trained on dimensional \gls{SER} on the same data by \citet{Wagner23-DOT}, and may therefore exhibit better performance due to this cascaded \gls{IID} finetuning.
The second-best model was \hlarge, with a \gls{UAR} of $.570$.

\begin{figure}[t]
    \centering
    \includegraphics[width=\columnwidth]{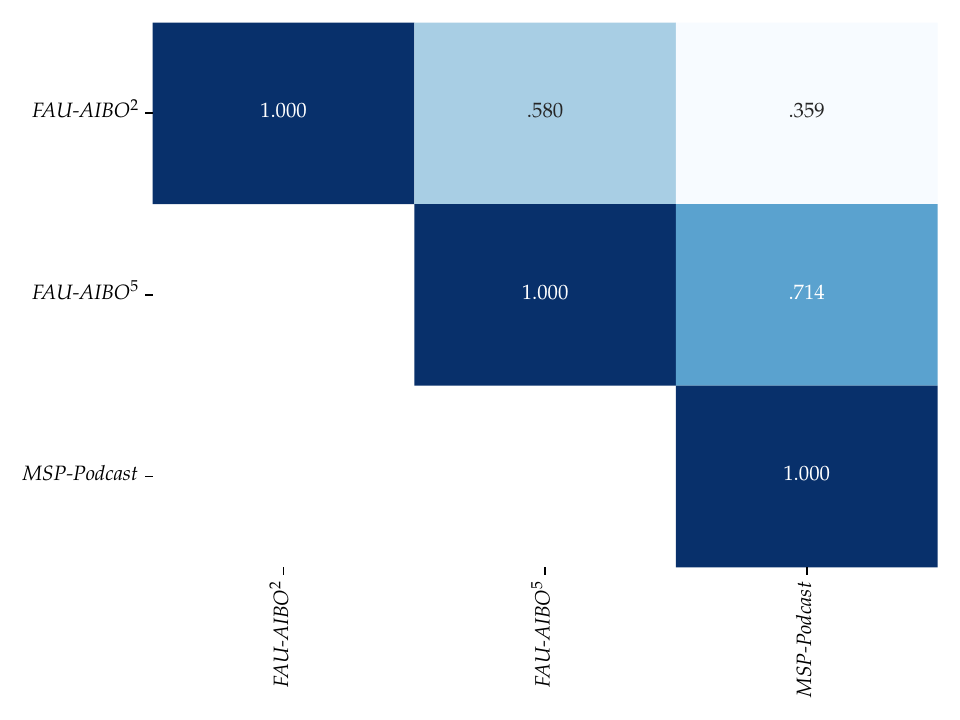}
    \caption{
    Pairwise agreement in the relative ranking of models
    measured as Spearman's $\rho$ between the \gls{UAR} achieved by each model
    for two different tasks.
    A higher agreement indicates
    that models which perform well for the one task
    will perform better for the other task as well.
    }
    \label{fig:model-ranks}
\end{figure}

\noindent
\textbf{Is model ranking consistent across tasks?} We considered model ranking across the seven different tasks, where we computed the pairwise Spearman's $\rho$ computed over model performance on each pair of tasks.
The results are presented in \cref{fig:model-ranks}.
We observed moderate-to-high correlations between different tasks.
Notably, performance on {\aibo}$^5$ correlated more with {\podcasteleven} ($.714$) than with {\aibo}$^2$ ($.580$), despite the fact that the latter two tasks share identical data.
This indicates that there is no `one-winner-takes-all' model.
Rather, as is consistent with the `no-free-lunch' theorem~\citep{Shalev14-UML}, a different model can be more appropriate for a given dataset and task.

\noindent
\textbf{Do newer/larger models perform better?} \cref{tab:year-macs} displays the Spearman's $\rho$ between model performance and the year each model was created, its \glspl{MAC}, and its number of parameters.
We observed very low correlations overall, with the highest correlation between complexity (measured in \glspl{MAC}) achieved for {\aibo}$^5$ being a meagre $.230$.
We acknowledge that these correlations heavily depend on our particular selection of models.
We account for this by also showing $95\%$ bootstrap \glspl{CI}.
The large variance in our results, ranging from low negative correlations ($>-.3$) to moderately high correlations ($>.3$), does not allow for robust conclusions.


\begin{table}[t]
    \centering
    \caption{
    Spearman's $\rho$ between model performance and year of publication, \glspl{MAC}, and number (\#) of parameters of each model for each dataset and task.
    We additionally show $95\%$ \glspl{CI} obtained using bootstrapping ($1000$ random samples with replacement).
    Note that model \glspl{MAC} and parameters do not account for feature extraction.
    }
    \label{tab:year-macs}
    \resizebox{\columnwidth}{!}{
   \begin{tabular}{lrrr}
    \toprule
    & \aibo$^2$ & \aibo$^5$ & \podcasteleven \\
    \midrule
    \textbf{Year}          & .122 [-.204 | .414] & .093 [-.201 | .384] & .050 [-.291 | .392]\\
    \textbf{\glspl{MAC}}   & .151 [-.157 | .446] & .230 [-.087 | .512] & .095 [-.263 | .416]\\
    \textbf{\# Parameters} & -.083 [-.353 | .195] & .085 [-.237 | .393] & .026 [-.319 | .356]\\
    \bottomrule
    \end{tabular}}
\end{table}

\noindent
\textbf{Agreement between different models:}
We first considered model agreement on {\aibo}.
The average pairwise agreement (percentage of instances where two models agree) for all models of the exploration phase was $70\%$ and $55\%$ for the $2$- and $5$-class models of {\aibo}, which rises to $80\%$ and $57\%$, respectively, when considering only the top-5 ones.
Overall, this demonstrates that models trained on similar data do not converge to an identical solution -- a finding congruent with the literature on underspecification~\citep{Damour22-UPC}.
Accordingly, the pairwise agreement for {\podcasteleven} was $55\%$/$79\%$ for all models and for the top-5 models respectively.
These numbers indicate a moderate-to-high agreement on model predictions,
especially for the best-performing models.
This illustrates 
that, despite model size or year of publication, 
models trained on the same data tend to learn similar -- but not identical -- concepts.

\noindent
\textbf{Model vs human performance:}
We also investigated whether samples that are harder to classify for humans are also harder for models.
The standard {\aibo} release comes with annotator confidences per instance, computed by taking the percentage of annotators who agree with the gold standard;
we thus defined `difficulty' as $1$ minus that confidence.
We then make the following observations when considering all models of the tuning phase:

a) We first adopted a model-agnostic measure of difficulty, which we defined as the number of models who disagree with the max-vote computed by all models on each instance -- this is akin to the computation of difficulty for the human annotators.
Spearman's $\rho$ between this measure and annotator disagreement was moderate ($.33$ and $.20$ for the $2$- and $5$-class problems).

b) We then adopted a model-specific measure of difficulty, defined as the standard cross-entropy loss for each instance, similar to \citet{Hacohen19-OTP} (see \citet{Rampp24-DTD} for our own investigations on the matter).
Different models have different rankings of instance difficulty, with average pairwise $\rho$ being $.51$ and $.33$ for the $2-$ and $5-$class problems.

c) Finally, we computed the Spearman's $\rho$ between each model's \gls{UAR} and its $\rho$ with annotator disagreement;
here, $\rho$ was $-.38/-.07$ for the $2$/$5$-class problem, indicating that larger agreement with human annotators did not lead to better performance (rather, the opposite).

We performed a similar analysis for {\podcasteleven}.
Here, we did not directly have annotator confidence, as samples where annotators disagree have been filtered out by the original authors\footnote{These were marked as ``no agreement'' in the original data and were excluded by filtering for the $4$ classes we use here.}.
Instead, we adopted two of the dimensional annotations (\emph{arousal} and \emph{valence}) as proxies for the difficulty of a sample.
Our intuition was that samples with a very high, or very low, arousal and/or valence, would be easier to distinguish than others.
Our model-agnostic measure of difficulty (percentage of models
which agree with the max-vote) had a Spearman's $\rho$ of $-.15$ with arousal and $-.07$ with valence.
The average Spearman's $\rho$ of the instance-level loss of each model with arousal was $.17$ and with valence $-.18$

Collectively, our results indicate that models appear to learn differently than humans, with small agreement to what constitutes an easy or hard example.
This was true for both {\aibo} and {\podcasteleven}.
This points to the fact that, on the one hand, there is still room for further growth, but also that models may fail in ways that are considered `catastrophic' by human observers.

\begin{table}[t]
    \centering
    \caption{
    \gls{UAR} results for best-performing architectures in the \emph{tuning phase} of \aibo.
    Showing best results obtained after tuning hyperparameters (batch size, optimiser, learning rate) and keeping the best-performing combination on the official test set.
    Also including $95\%$ \glspl{CI} for our models computed with bootstrapping.
    State-of-the-art results taken from original works.
    }
    \label{tab:tuning:aibo}
\begin{tabular}{lcc}
\toprule
       \textbf{Model} &   {\aibo}$^2$ &   {\aibo}$^5$ \\
\midrule
    2009 Baseline & .677 & .382 \\
    2009 Winners  & .703 & .417 \\
    2009 Fusion   & .712 & .440 \\
    \citet{Zhao19-EDS} & N/A & .454  \\
    \midrule
   \emph{IS10-d-lstm} & .685 [.674 - .696] &   .394 [.377 - .411] \\
   \emph{Resnet50} & .690 [.680 - .701] &   .423 [.405 - .441] \\
   \cnn & .672 [.661 - .683] &   .448 [.428 - .467] \\
   \wemo & \textbf{.717 [.706 - .728]} &   .448 [.431 - .465] \\
   \emph{Whisper}$^t$ & .707 [.696 - .718] & \textbf{.454 [.437 - .472]} \\
\bottomrule
\end{tabular}
\end{table}

\begin{table}[t]
    \centering
    \caption{
    \gls{UAR} results for best-performing architectures in the \emph{tuning phase} of \podcasteleven.
    Showing best results obtained after tuning hyperparameters (batch size, optimiser, learning rate) and keeping the best-performing combination on the official test set.
    Also including $95\%$ \glspl{CI} for our models computed with bootstrapping.
    }
    \label{tab:tuning:msp}
    \begin{tabular}{c|c}
    \toprule
    \textbf{Model} & \gls{UAR}\\
    \midrule
    {\effnet} & .531 [.523 - .540]\\
    \hlarge & .624 [.615 - .633]\\
    \wrobust & .563 [.554 - .571]\\
    \wvox & .554 [.546 - .562]\\
    \wemo & .650 [.642 - .658]\\
    \bottomrule
    \end{tabular}
\end{table}

\begin{figure*}[t]
    \centering
    \includegraphics[width=0.33\textwidth]{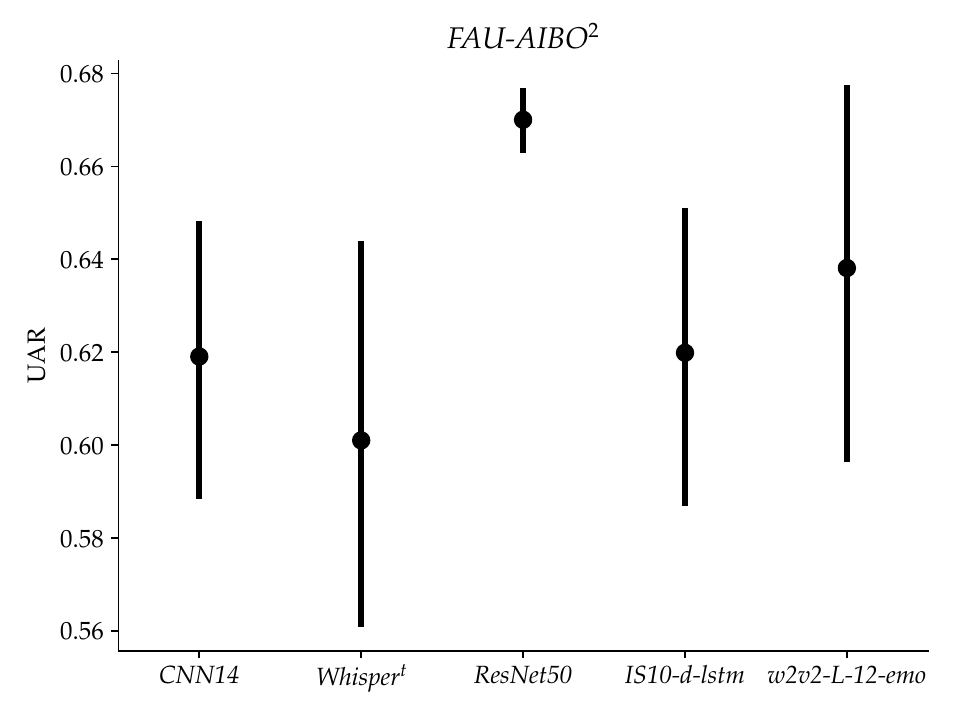}~%
    \includegraphics[width=0.33\textwidth]{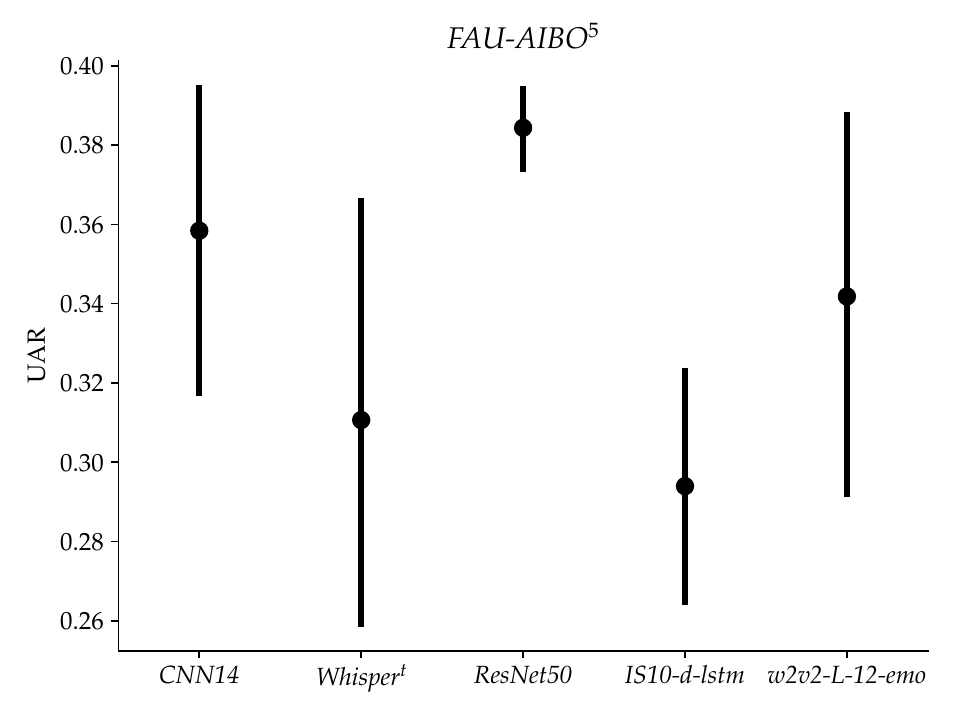}~%
    \includegraphics[width=0.33\textwidth]{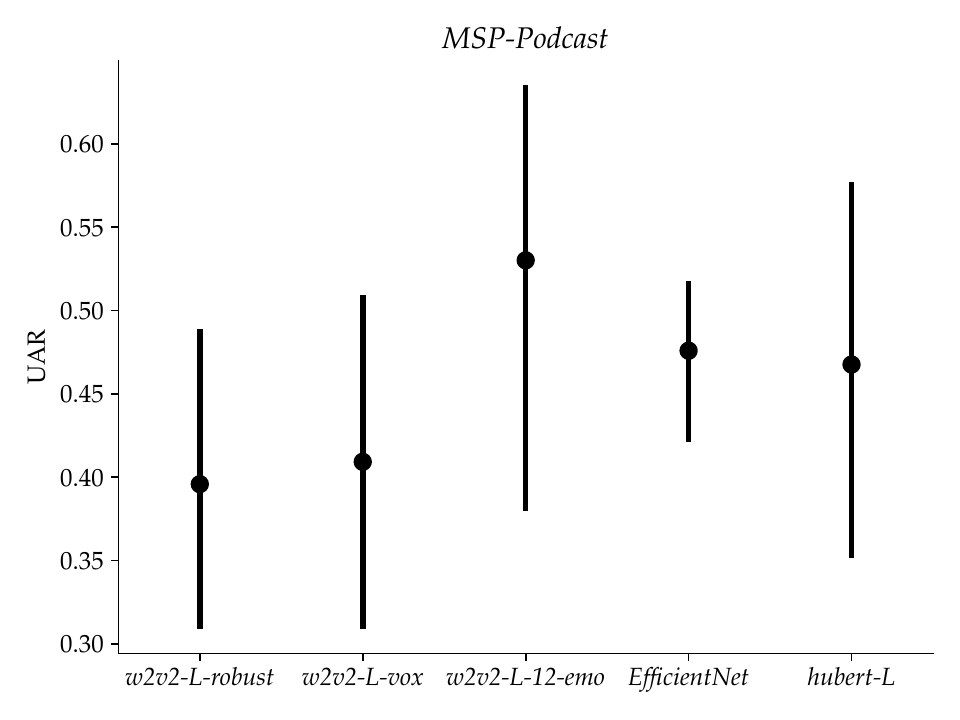}
    \caption{
    Results of the \emph{tuning phase} for {\aibo}$^2$ (left), {\aibo}$^5$ (middle), and {\podcasteleven} (right).
    Showing distribution of \gls{UAR} values for the different hyperparameters tested
    for the top-5 performing models in each task.
    }
    \label{fig:results:tuning}
\end{figure*}


\noindent
\textbf{Tuning phase:} Turning to the \emph{tuning phase} of {\aibo} and {\podcasteleven}, \cref{tab:tuning:aibo} and \cref{tab:tuning:msp} show the \emph{best} performance for the top five models
from \cref{tab:exploration} after optimising standard hyperparameters (optimiser, learning rate, batch size).
For {\aibo}, {\wemo} yielded the best performance for the $2$-class problem of {\aibo} ($.717$) and \emph{Whisper}$^t$ for its $5$-class problem ($.454$) -- in both cases, we reached results better than the challenge winners, albeit with the gains for the $2$-class problem being marginal.
Given that \emph{Whisper} has been trained for multilingual \gls{ASR} (including German), and that performance improvements on valence prediction for English speech heavily depend on implicit linguistic knowledge (see \citet{Triantafyllopoulos22-PSE}), we expect \emph{Whisper}'s success to be also attributed to that aspect (see \cref{ssec:text} for how text-based models performed).
However, it is still the case that all models we tested remained close to or even below the original challenge baseline and winners, and especially the fusion of the top challenge submissions.
This remained so even after selecting only the best-performing model out of all tested hyperparameters, essentially following a generally erroneous practice of overfitting.
This was done intentionally to gauge performance under the most optimistic of settings -- that of virtually unrestricted evaluation runs.
We note that the original challenge participants were given $25$ runs each.
This shows how the gains obtained here must be further tempered to account for more runs on our side.

For {\podcasteleven}, the best performing model was a variant of {\wemo}, trained with \emph{SGD}, a learning rate of $.01$, and a batch size of $4$, which reached a \gls{UAR} of $.650$ $[.642 - .658]$ on the test set.
We release this model, which we hence refer to as {\wser} publicly\footnote{\url{https://huggingface.co/autrainer/msp-podcast-emo-class-big4-w2v2-l-emo}}.
The performance for other models also improved, with the best hyperparameter combination for {\hlarge} reaching $.624 [.615 - .633]$, illustrating once more how hyperparameter tuning can make a substantial difference in performance comparisons.

\cref{fig:results:tuning} shows all tuning results on both {\aibo} tasks and {\podcasteleven}.
We observed a very high variability when changing hyperparameters -- a well-known phenomenon impacting \gls{DL}~\citep{Perrone18-SHT}.
On the positive side, this tuning allowed us to obtain much higher performance -- on {\podcasteleven} this gave rise to a \gls{UAR} of $.650$ for {\wemo}.
However, the high variability calls attention to the fact that a different choice of hyperparameters in the exploration phase could have resulted in both different overall performance and, crucially, a different ranking across architectures.
Unfortunately this is a problem that plagues all \gls{DL} research at the moment, with the only possible solution being the investment in more compute power to explore a wider space of hyperparameters (which was not possible in our case).

\subsubsection{OOD results}

\begin{table}[t!]
    \centering
    \caption{
    \gls{OOD} \gls{UAR} results for the $4$-class categorical \gls{SER} models trained on {\podcasteleven} when evaluated on {\emodb} and {\iemocap}.
    }
    \label{tab:msp:ood}
    \begin{tabular}{lrr}
    \toprule
    & \emodb & \iemocap \\
    \midrule
    \emph{IS09-s-mlp} & \cellcolor{w0}{.250} & \cellcolor{w0}{.250} \\
    \emph{IS09-d-lstm} & \cellcolor{w3}{.487} & \cellcolor{w0}{.339} \\
    \emph{IS10-d-lstm} & \cellcolor{w0}{.423} & \cellcolor{w0}{.354} \\
    \emph{IS10-s-mlp} & \cellcolor{w15}{.553} & \cellcolor{w25}{.489} \\
    \emph{IS11-d-lstm} & \cellcolor{w0}{.440} & \cellcolor{w0}{.373} \\
    \emph{IS11-s-mlp} & \cellcolor{w28}{.630} & \cellcolor{w29}{.503} \\
    \emph{AlexNet} & \cellcolor{w0}{.250} & \cellcolor{w0}{.250} \\
    \emph{IS12-d-lstm} & \cellcolor{w6}{.504} & \cellcolor{w0}{.315} \\
    \emph{IS12-s-mlp} & \cellcolor{w23}{.598} & \cellcolor{w25}{.489} \\
    \emph{IS13-s-mlp} & \cellcolor{w28}{.627} & \cellcolor{w27}{.495} \\
    \emph{IS13-d-lstm} & \cellcolor{w3}{.489} & \cellcolor{w0}{.365} \\
    \emph{eGeMAPS-s-mlp} & \cellcolor{w10}{.530} & \cellcolor{w26}{.491} \\
    \emph{eGeMAPS-d-lstm} & \cellcolor{w0}{.361} & \cellcolor{w0}{.382} \\
    \emph{Resnet50} & \cellcolor{w39}{.691} & \cellcolor{w20}{.470} \\
    \emph{VGG}$^{13}$ & \cellcolor{w0}{.250} & \cellcolor{w0}{.250} \\
    \emph{VGG}$^{19}$ & \cellcolor{w0}{.250} & \cellcolor{w0}{.250} \\
    \emph{VGG}$^{11}$ & \cellcolor{w14}{.549} & \cellcolor{w3}{.409} \\
    \emph{VGG}$^{16}$ & \cellcolor{w22}{.596} & \cellcolor{w25}{.490} \\
    \emph{IS16-s-lstm} & \cellcolor{w32}{.649} & \cellcolor{w19}{.466} \\
    \emph{IS16-d-mlp} & \cellcolor{w0}{.460} & \cellcolor{w0}{.351} \\
    \emph{CRNN}$^{18}$ & \cellcolor{w32}{.648} & \cellcolor{w13}{.445} \\
    \emph{CRNN}$^{19}$ & \cellcolor{w31}{.642} & \cellcolor{w33}{.519} \\
    \effnet & \cellcolor{w21}{.590} & \cellcolor{w2}{.406} \\
    \wlarge & \cellcolor{w0}{.250} & \cellcolor{w0}{.250} \\
    \wbase & \cellcolor{w0}{.250} & \cellcolor{w0}{.250} \\
    \emph{ConvNeXt}$^b$ & \cellcolor{w32}{.653} & \cellcolor{w14}{.447} \\
    \emph{ConvNeXt}$^l$ & \cellcolor{w38}{.684} & \cellcolor{w30}{.505} \\
    \emph{ConvNeXt}$^t$ & \cellcolor{w42}{.705} & \cellcolor{w31}{.510} \\
    \cnn & \cellcolor{w44}{.715} & \cellcolor{w9}{.429} \\
    \emph{ETDNN} & \cellcolor{w23}{.600} & \cellcolor{w28}{.501} \\
    \emph{ConvNeXt}$^s$ & \cellcolor{w35}{.669} & \cellcolor{w33}{.519} \\
    \hbase & \cellcolor{w0}{.250} & \cellcolor{w0}{.250} \\
    \emph{Swin}$^b$ & \cellcolor{w0}{.250} & \cellcolor{w0}{.250} \\
    \emph{Swin}$^s$ & \cellcolor{w0}{.250} & \cellcolor{w0}{.250} \\
    \emph{Swin}$^t$ & \cellcolor{w0}{.250} & \cellcolor{w0}{.320} \\
    \emph{AST} & \cellcolor{w6}{.505} & \cellcolor{w9}{.431} \\
    \wrobust & \cellcolor{w54}{.771} & \cellcolor{w28}{.499} \\
    \wvox & \cellcolor{w56}{.783} & \cellcolor{w35}{.526} \\
    \hlarge & \cellcolor{w51}{.756} & \cellcolor{w40}{.544} \\
    \emph{Whisper}$^b$ & \cellcolor{w0}{.250} & \cellcolor{w0}{.250} \\
    \emph{Whisper}$^t$ & \cellcolor{w0}{.401} & \cellcolor{w0}{.377} \\
    \emph{Whisper}$^s$ & \cellcolor{w6}{.503} & \cellcolor{w0}{.381} \\
    \wemo & \cellcolor{w60}{.806} & \cellcolor{w60}{.617} \\
    \bottomrule
    \end{tabular}
\end{table}

\gls{OOD} results are shown in \cref{tab:msp:ood}.
In many cases, \gls{OOD} \gls{UAR} is, surprisingly, higher than \gls{IID}.
We interpret this a side-effect of the datasets being much `easier' than {\podcasteleven}, as they only contain acted, and very prototypical, data.
Nevertheless, it is a promising sign of model generalisation.
The best \gls{OOD} performance in both cases was achieved by {\wemo}, which reached a \gls{UAR} of $.806$ on {\emodb} and $.617$ on {\iemocap}.
The Spearman's $\rho$ of \gls{IID} with \gls{OOD} performance was $.909$ for {\emodb} and $.843$ for {\iemocap}.
This is another positive finding as it entails that selecting the best-performing model based on \gls{IID} performance alone will translate to the best-performing model on \gls{OOD} as well.
However, the Spearman's $\rho$ between year, \glspl{MAC}, or number of parameters and \gls{OOD} \gls{UAR} remained low for both {\emodb} and {\iemocap} (year: $r=.112/.098$; \glspl{MAC}: $r=.166/.064$; $\#$ parameters: $.172/.096$).

\subsubsection{Transfer learning dynamics of categorical SER models}
We next considered
the transfer learning dynamics
of categorical \gls{SER} models
trained on {\podcasteleven},
as this is the largest dataset
we have trained on.
We computed the \gls{CKA}~\citep{Kornblith19-SON},
a measure of similarity for hidden representations
using {\emodb} as a probing dataset
due to its smaller size.
We did so for all transformer models,
namely,
{\wbase}, {\wlarge}, {\wvox}, {\wrobust}, {\wemo},
{\hbase}, and {\hlarge},
as well as two exemplary \glspl{CNN},
{\cnn} and {\effnet}.
For each model,
we computed the \gls{CKA}
between hidden representations
extracted from the initial state
before fine-tuning,
and all subsequent intermediate states
which were saved every second epoch.
For transformers,
we extracted the representations
after each hidden transformer block,
for {\cnn} we did so after each of its convolutional blocks,
and similarly for {\effnet}.

\begin{figure*}[t!]
    \centering
    \begin{subfigure}[t]{.3\textwidth}
        \centering\includegraphics[width=\textwidth]{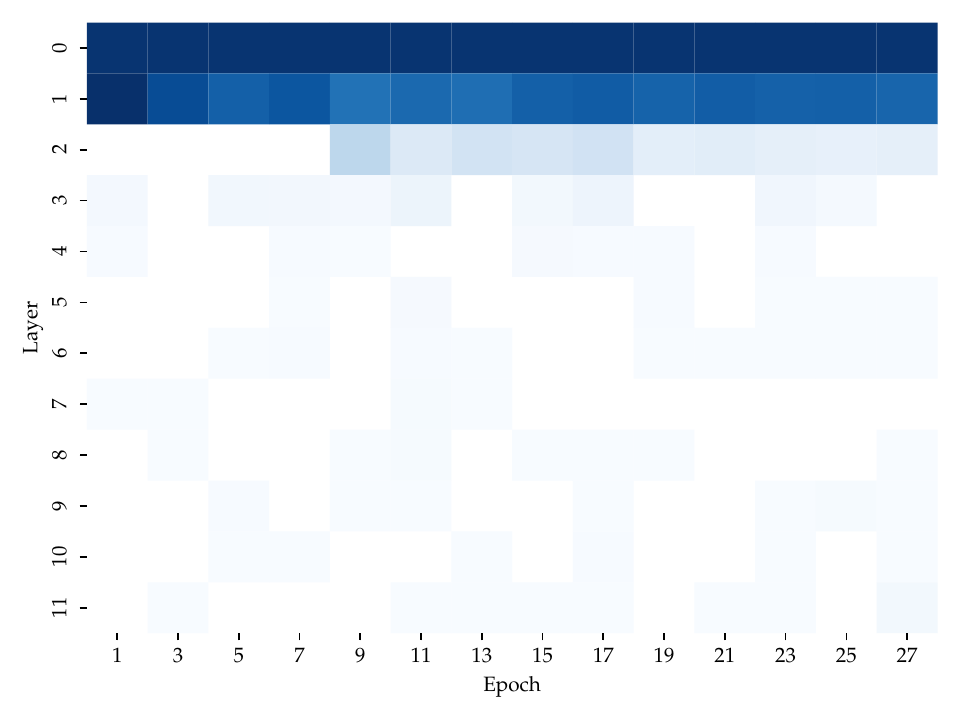}
        \caption{\wbase}
    \end{subfigure}
    \begin{subfigure}[t]{.3\textwidth}
        \centering\includegraphics[width=\textwidth]{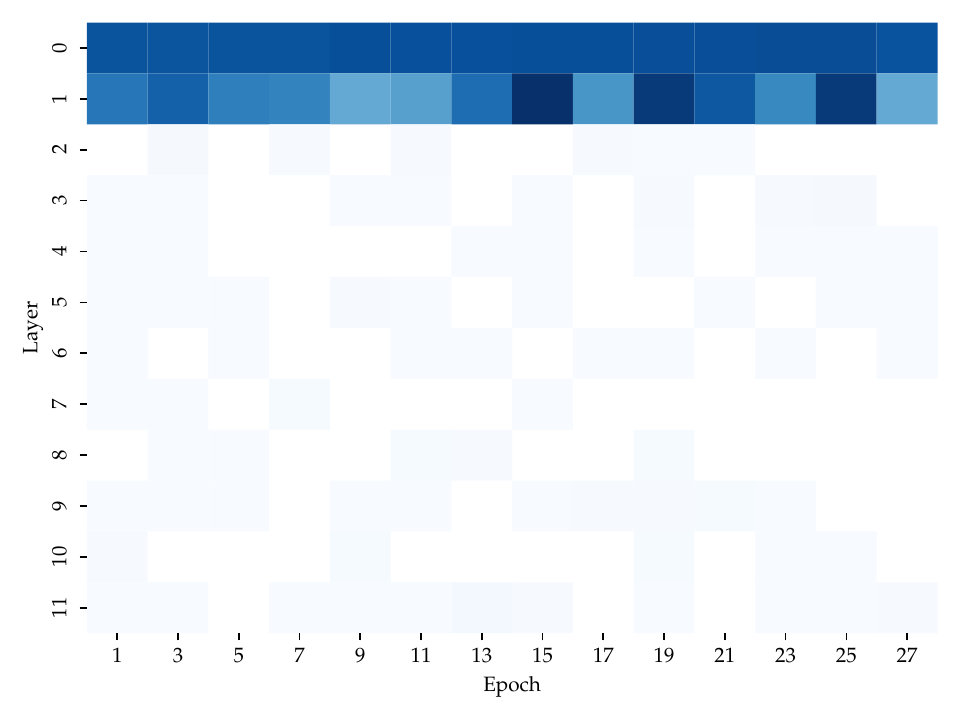}
        \caption{\hbase}
    \end{subfigure}
    \begin{subfigure}[t]{.3\textwidth}
        \centering\includegraphics[width=\textwidth]{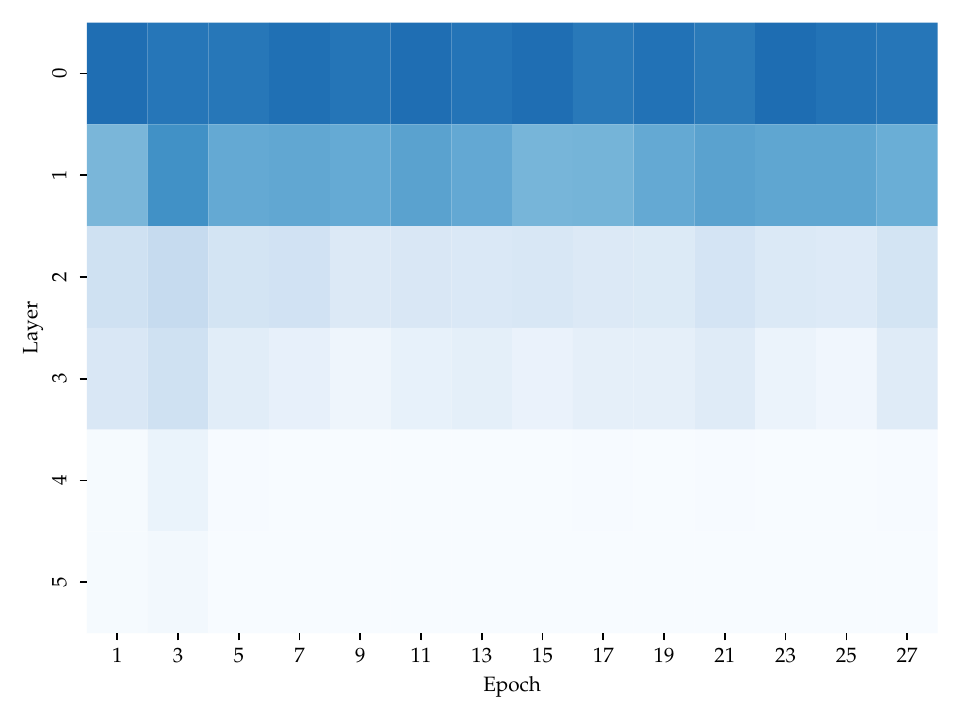}
        \caption{\cnn}
    \end{subfigure}
    \begin{subfigure}[t]{.3\textwidth}
        \centering\includegraphics[width=\textwidth]{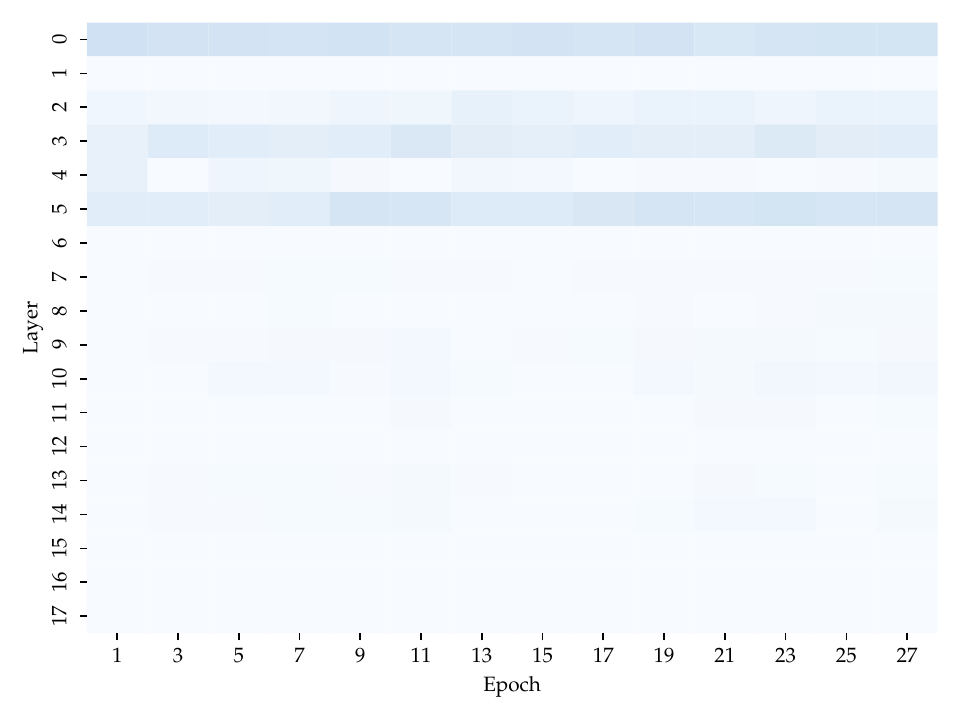}
        \caption{\effnet}
    \end{subfigure}
    \begin{subfigure}[t]{.3\textwidth}
        \centering\includegraphics[width=\textwidth]{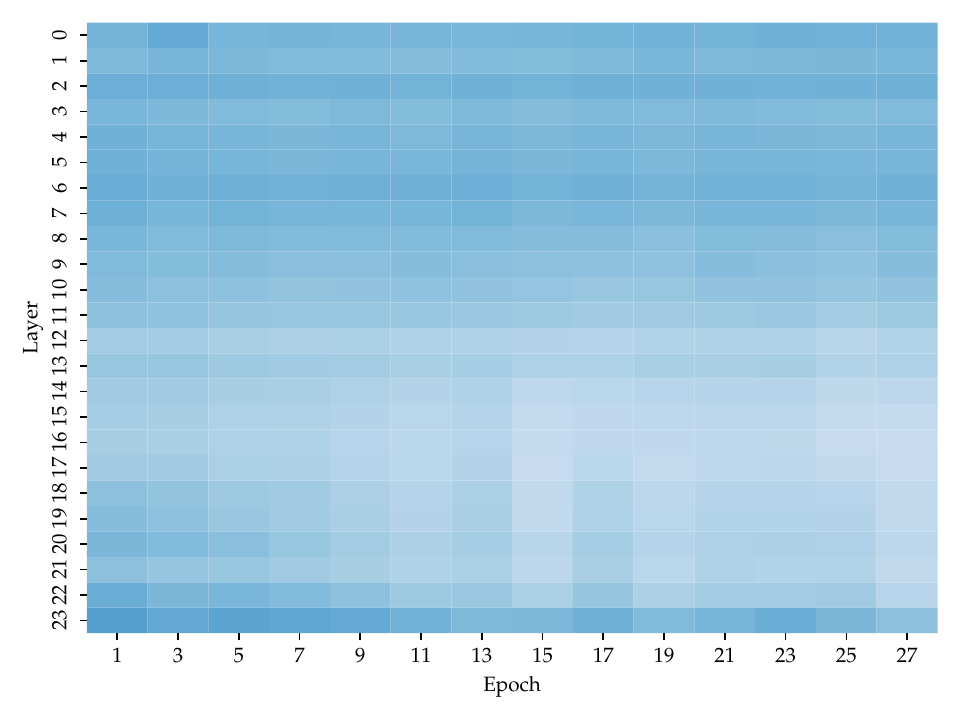}
        \caption{\hlarge}
    \end{subfigure}
    \begin{subfigure}[t]{.3\textwidth}
        \centering\includegraphics[width=\textwidth]{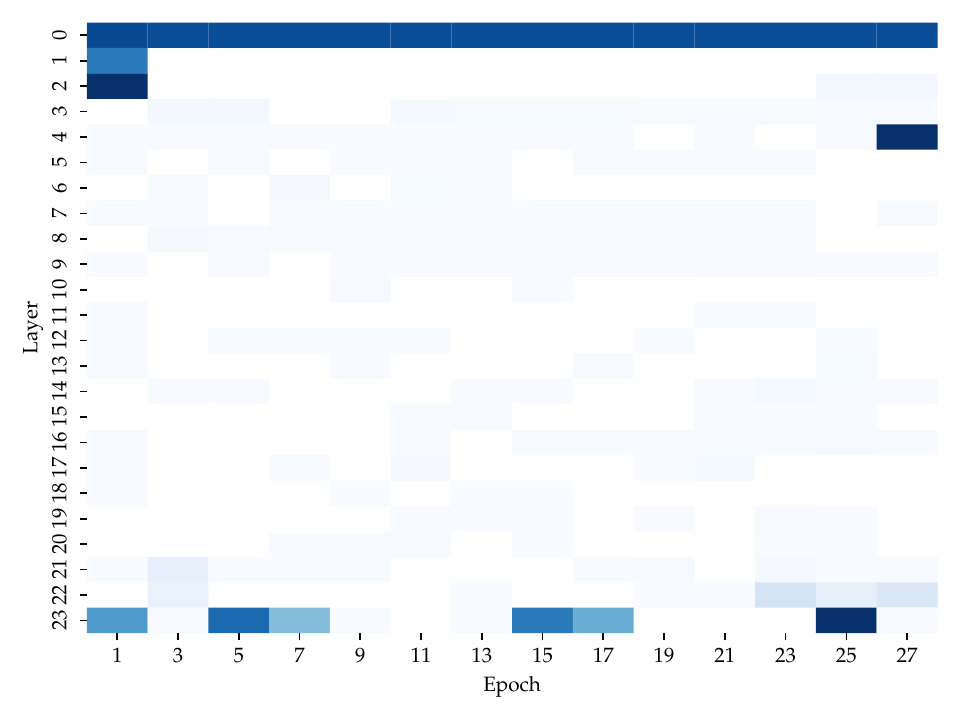}
        \caption{\wlarge}
    \end{subfigure}
    
    \begin{subfigure}[t]{.3\textwidth}
        \centering\includegraphics[width=\textwidth]{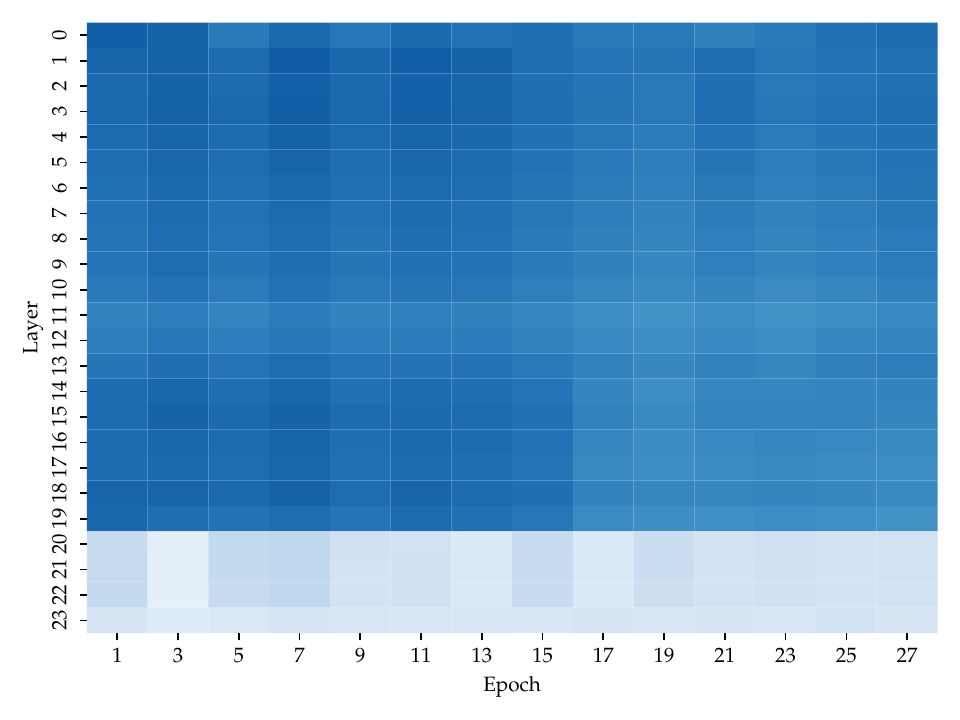}
        \caption{\wrobust}
    \end{subfigure}
    \begin{subfigure}[t]{.3\textwidth}
        \centering\includegraphics[width=\textwidth]{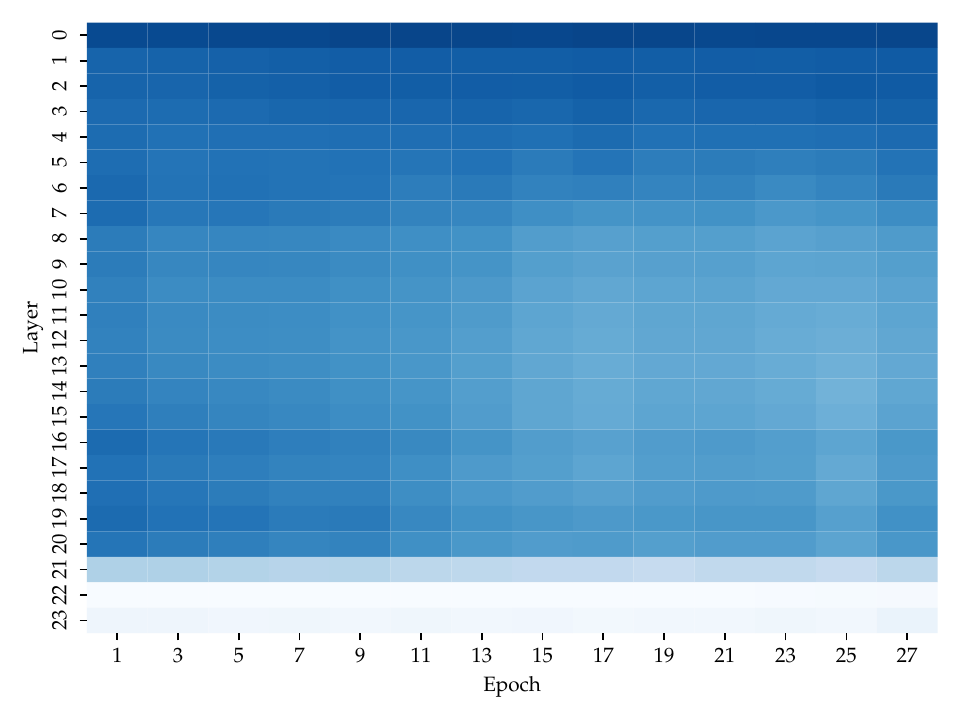}
        \caption{\wvox}
    \end{subfigure}
    \begin{subfigure}[t]{.3\textwidth}
        \centering\includegraphics[width=\textwidth]{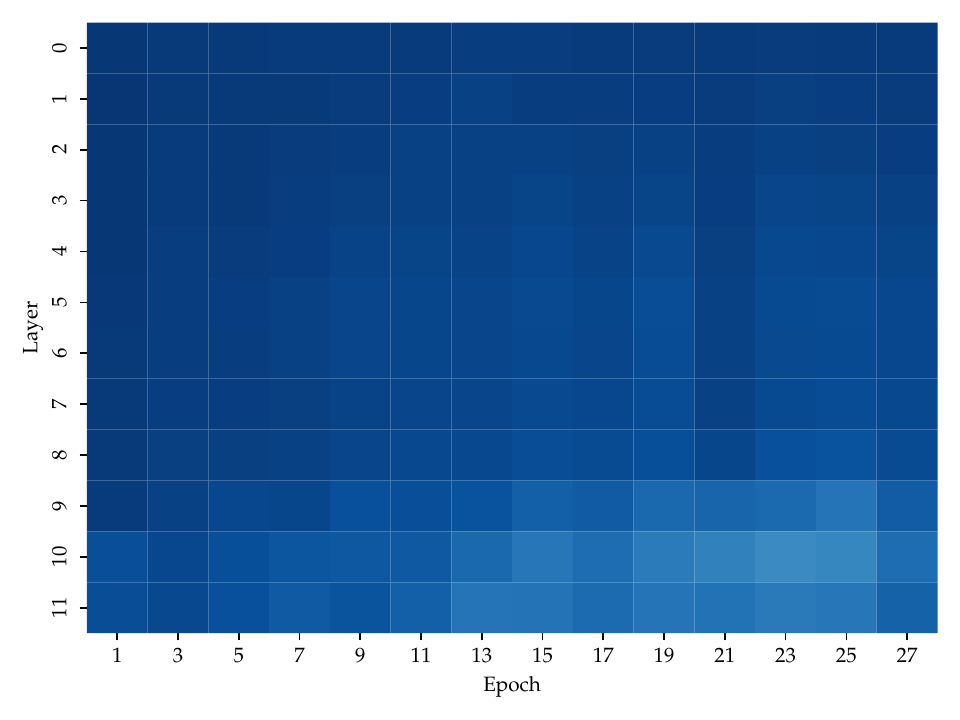}
        \caption{\wemo}
    \end{subfigure}
    \caption{
    Transfer learning dynamics of models fine-tuned on $4$-class \gls{SER} using {\podcasteleven}.
    Plots show \gls{CKA} between representations extracted at the initial state
    and at intermediate checkpoints stored every second epoch.
    A higher \gls{CKA} (darker colour) corresponds to higher similarity
    between the two states, {\ie} a higher amount of feature reuse.
    }
    \label{fig:replication:dynamics}
\end{figure*}

Results are shown in \cref{fig:replication:dynamics}.
Broadly,
we note that all models 
exhibited a pattern
of changing bottom-first (closer to the output),
consistent both with the intuition
that earlier layers learnt more general,
and thus re-usable features~\citep{Bengio12-DLO},
and prior results~\citep{Palanisamy20-RCM, Neyshabur20-WIB}.
However,
a closer look at individual models
reveals important differences.
Transformers
that saw more adaptation 
of earlier layers,
like {\wbase}, {\hbase}, and {\wlarge},
performed at chance-level performance.
This is an indication
that deviating a lot from the pretrained state
can have catastrophic effects.
We hypothesise
that this was a side-effect
of using a learning rate
that was too high for these models.
Interestingly,
a similar pattern was exhibited by {\effnet};
however,
this model still achieved good \gls{SER} performance.
We hypothesise
that this particular model
needed to adjust its representations more
as it had been only pretrained on \emph{ImageNet} data.
Lastly,
{\wemo} showed the least adaptation overall,
which is expected
as the model had already been trained
on {\podcastseven} in a previous step
(for dimensional \gls{SER})
and had thus already adapted to
the distribution of the data.

\subsubsection{Probing transformer representations}

Results are shown in \cref{tab:replication:probing}.
We note that better-performing models
showed a higher Pearson's $r$ overall
across all features,
indicating that
representations that contained more information
about those features
led to a better \gls{SER} performance.
Comparing errors across features
also uncovers interesting insights.
For instance,
$\mu$(P) had a much higher Pearson's $r$
than $\sigma$(P),
which shows that transformers
were better at grasping
average pitch
rather than pitch variations
as also shown by the high errors on jitter.
Given the importance 
of pitch variations
for \gls{SER}~\citep{Banse96-API},
this might present an avenue for future improvements.

\begin{table*}[t]
    \centering
    \caption{
    Average Pearson's $r$ over all layers and epochs when probing transformer representations for their knowledge of acoustic features using linear probes on {\emodb}.
    A higher Pearson's $r$ -- either positive or negative -- reflects the strength with which this feature is encoded in internal representations, and how it might be possibly utilised during inference.
    The last column shows \gls{UAR} taken from \cref{tab:exploration}.
    }
    \label{tab:replication:probing}
    \resizebox{\textwidth}{!}{
    \begin{tabular}{c|cccccccccc|c}
    \toprule
        \textbf{Model} & $\mu$(P) & $\sigma$(P) & $\mu$(L) & $\mu$(L)\,[dB] & $\mu$(J) & $\mu$(S) & $\mu$(HNR) & $\mu$(F1) & $\mu$(F2) & $\mu$(F3) & \gls{UAR}\\
    \midrule
        \wbase    & .863 & .172 & .454 & .271 & .185 & .318 & .519 & .603 & .458 & .313 & .250\\
        \hbase    & .523 & .149 & .304 & .207 & .055 & .114 & .250 & .380 & .306 & .228 & .250\\
        \wlarge   & .547 & .187 & .292 & .200 & .060 & .142 & .261 & .365 & .286 & .217 & .250\\
        \wrobust  & .926 & .088 & .361 & .297 & .112 & .247 & .518 & .726 & .597 & .409 & .555\\
        \wvox     & .941 & .213 & .504 & .435 & .175 & .275 & .584 & .772 & .653 & .473 & .561\\
        \hlarge   & .944 & .220 & .451 & .382 & .100 & .272 & .547 & .781 & .669 & .471 & .570\\
        \wemo     & .959 & .237 & .533 & .420 & .188 & .350 & .649 & .815 & .697 & .521 & .609\\
    \bottomrule
    \end{tabular}}
\end{table*}


\subsubsection{Robustness to additive noise}
The Spearman's $\rho$
between the \gls{UAR}
achieved for clean and noisy audio
was $.953$.
However,
as before,
the Spearman's $\rho$ 
between noisy \gls{UAR}
and year of publication ($.05$),
\glspl{MAC} ($.10$),
and $\#$ of parameters ($.03$)
was extremely low.
Thus,
while \gls{IID} performance
showed a strong correlation with robustness,
this robustness did not increase
based on the year of publication
or model size.
Detailed noise robustness results are shown
in \cref{app:robustness}
where we observe
substantial degradations
of model performance
even for the newest models,
with the best-performing {\wemo}
dropping from a \gls{UAR} of $.609$
to one of $.546$.


\subsubsection{Individual fairness}
We finish this section 
with a discussion of 
speaker-level performance
for audio-based models.
To do so,
we computed the speaker-level performance
for each task
and used that as the utility to compute the Gini index.

We are interested in two main questions:
a) what was the average \emph{equality} 
observed for a dataset
across models,
as measured by the Gini index, and,
b) were models that performed better
on the instance-level
also more fair
across speakers,
{\ie} what was the Spearman's $\rho$ 
between the \gls{UAR} of a model on the instance-level
and its Gini index across speakers?
We only considered models
that performed better than chance,
defined as a \gls{UAR} larger than $.01$ more
than the corresponding random-chance \gls{UAR},
{\ie} $.51$ for $2$-class, 
$.26$ for $4$-class,
and $.21$ for $5-$class,
respectively.
Our results are presented in \cref{tab:pers:gini}.

\begin{table}[t]
    \centering
    \caption{
    Average Gini index
    of speaker-level performance
    for all models
    and Spearman's $\rho$ between that Gini index
    and global performance (\gls{UAR}).
    A higher $\mu$(Gini) indicates more inequality
    across speakers for all models trained on a given task.
    A lower $\rho$ shows that models with the highest performance
    have a lower inequality as well. 
    }
    \label{tab:pers:gini}
    \begin{tabular}{c|cc}
    \toprule
    \textbf{Task} & $\rho$ & $\mu$(Gini)\\
    \midrule
        \aibo$^2$ & .075 & .078 \\
        \aibo$^5$ & -.015 & .139  \\
        \podcasteleven & -.344 & .329  \\
    \bottomrule
    \end{tabular}
\end{table}

We note that the only task 
which showed a moderate average inequality across speakers
is {\podcasteleven}, with an average Gini index of $.329$.
On the positive side,
it also showed a high \emph{negative} correlation
between instance-level \gls{UAR} 
and speaker-level Gini index ($-.344$).
This means that models 
which performed better overall
across all instances
were also \emph{more fair}
towards different speakers.
The {\aibo} tasks exhibited a low to moderate average Gini index
and a negligible correlation between instance-level \gls{UAR}
and the Gini index across speakers.
Based on this,
we conclude that speaker inequality
affects different datasets and tasks to different degrees,
but it remains an open avenue for future research
considering the potential of \emph{personalisation} methods
to improve fairness and overall performance~\citep{Triantafyllopoulos24-EPF}.

The Spearman's $\rho$
between the Gini index
and year of publication ($-.04$),
\glspl{MAC} ($.-25$),
and $\#$ of parameters ($-.07$)
for {\podcasteleven} 
was either zero
or mildly negative
showing that
bigger models
were slightly
more fair
than smaller ones.
However,
this trend is contradicted
by results for {\aibo}$^2$ ($.29/.55/.53$)
and {\aibo}$^5$ ($.18/.32/.26$)
which showed moderate-to-strong
Spearman's $\rho$ between the Gini index
and year of publications, \glspl{MAC}, and $\#$ of parameters,
pointing towards \emph{lower individual fairness}
for newer and bigger models.



\subsection{Text-based models}
\label{ssec:text}

In addition,
we used state-of-the-art
text-based models
to predict emotion
on both {\aibo} tasks
and {\podcasteleven}.
\gls{UAR} results are shown in \cref{tab:replication:text}.

The highest performance 
for {\aibo}$^2$ 
was achieved by \emph{DistilBERT}
at $.667$,
for {\aibo}$^5$ by \emph{Llama-3}
at $.401$,
and for {\podcasteleven} by \emph{Llama-2}
at $.564$,
which was essentially on-par
with \emph{Llama-3} at $.563$
and \emph{Mistral} at $.560$.
Notably,
all these results were lower
than the best-performing
audio-based models
though competitive
with models trained on the exploration phase.
For instance,
\emph{Llama-2} results on {\podcasteleven}
would rank third in our \emph{exploration phase},
trailing marginally behind {\hlarge}
which scored at $.570$.
This demonstrates
how text contains vital information
for the \gls{SER} task of {\podcasteleven}.

Results on {\aibo}
were overall lower,
with text-based models
scoring in the middle of the `competition'.
We hypothesise
that this is a side-effect
of the fact that the textual content in {\aibo}
is much more limited than in {\podcasteleven}
given the Wizard-of-Oz scenario
in which it was recorded.
Nevertheless,
the fact that
text-based models
obtained better-than-chance
performance
is in itself remarkable,
given how limited the textual content is,
and indicates that
there are potential systematic biases
with respect to the linguistic content
of certain emotions.
We discuss 
the linguistic content 
of both datasets 
further
in \cref{app:linguistics}.

\begin{table}[!t]
    \centering
    \caption{
    \gls{UAR} results for text-based models on {\aibo} and {\podcasteleven}.
    }
    \label{tab:replication:text}
    \begin{tabular}{l|lll}
    \toprule
    \textbf{Model} & {\aibo}$^2$ & {\aibo}$^5$ & {\podcasteleven}\\
    \midrule
    {\bert}            & .647 & .393 & .528\\
    \emph{RoBERTa}     & .646 & .381 & .541\\
    \emph{DistilBERT}  & \textbf{.667} & .391 & .523\\
    \emph{Electra}     & .666 & .386 & .536 \\
    \emph{Llama-2}     & .662 & .389 & \textbf{.564} \\
    \emph{Llama-3}     & .645 & \textbf{.401} & .563 \\
    \emph{Mistral}     & .623 & .378 & .560 \\
    \bottomrule
    \end{tabular}
\end{table}

\subsubsection{Complementarity between audio \& text}
We finally considered
the complementarity
between audio-based
and text-based models.
In the present subsection,
we rely on error analysis
to investigate
whether text-based models,
which generally underperform
audio-based ones,
brought additional benefits
for classification
or merely predicted correctly
a subset of the data
where audio-based models
already performed well.

For {\podcasteleven}
we compared two exemplary
audio-based models
from the exploration phase,
{\wemo}, as the best-performing transformer,
and {\effnet}, as the best-performing non-transformer model,
and the best-performing text-based model,
\emph{Llama-2}.
We also considered {\aibo}$^2$
using {\wemo},
{\cnn},
and \emph{DistilBERT}
as the three models
showing a more balanced performance
over {\aibo}$^2$ and {\aibo}$^5$.

We first examined the pairwise agreement
between all model combinations,
measured as the percentage of samples
where two models
make the same decision.
For {\podcasteleven}, 
this lied at $.556$
between \emph{Llama-2} and {\wav},
$.451$ between \emph{Llama-2} and {\effnet},
and $.639$ between {\effnet} and {\wav}.
For {\aibo}$^2$,
we also observed a larger agreement
between {\wemo} and {\cnn} ($.815$)
than between {\wemo} and \emph{DistilBERT} ($.666$)
or {\cnn} and \emph{DistilBERT} ($.672$).

\begin{figure*}
    \includegraphics[width=.5\textwidth]{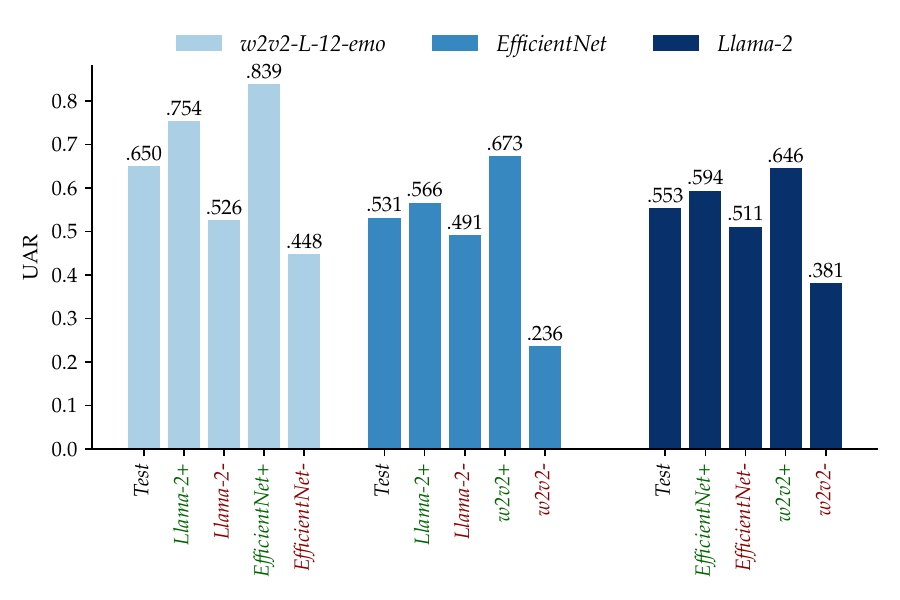}~%
    \includegraphics[width=.5\textwidth]{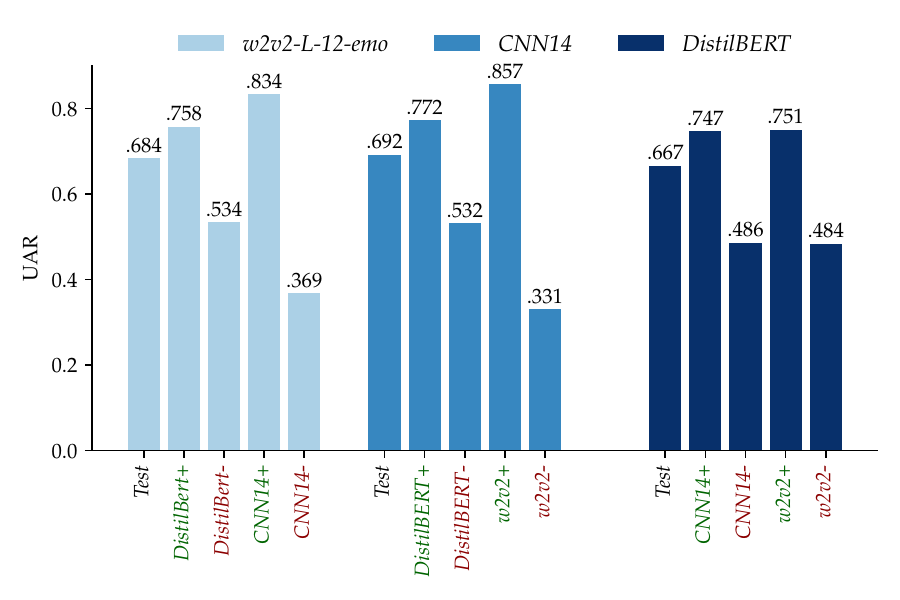}
    \label{fig:complementarity}
    \caption{
    \gls{UAR} for each model on {\podcasteleven} (left) and {\aibo}$^2$ (right) on the entire test set vs on the subset that each of the three models examined here classifies correctly ($+$; green) or incorrectly ($-$; red).
    }
\end{figure*}

Additionally,
we measured their complementarity
by comparing the \gls{UAR} of each model
on the entire test set (\emph{Test-1})
vs its \gls{UAR}
on the subset of the test set
that each other model classifies correctly ($+$; green)
or incorrectly ($-$; red).
These results are shown in \cref{fig:complementarity}.
On {\podcasteleven},
{\wemo} showed higher \gls{UAR}
when either \emph{Llama-2}
or \emph{EfficientNet}
performed well
-- a sign that its correct predictions
were a superset of the correct predictions
of the other two models.
In contrast, \emph{EfficientNet} and \emph{Llama-2}
only showed a performance improvement
when evaluated on the samples that {\wemo} 
classified correctly,
thus showing little overlap between these pairs of models.


These results
lead us to conclude
that the complementarity
was much higher
between {\effnet}
and \emph{Llama-2},
than between {\wemo}
and any of the other two models.
We hypothesise
that this happens because
{\wemo} already captured
linguistics implicitly,
as shown in \citet{Triantafyllopoulos22-PSE},
whereas {\effnet} did not.
Seen in a different light,
this means that {\wemo}
-- and, generally, other pretrained transformers --
seem capable of combining
`the best of both worlds'.


Overall,
this investigation revealed
that audio-based models
agreed more with one another
than they agreed
with a text-based model.
However,
the differences for {\aibo}
were less pronounced
than for {\podcasteleven}.
We hypothesise
that this was the case because
a) {\aibo} is a German dataset
and {\wemo} had not been pre-trained
on German data, 
and b) {\aibo} was collected
in a Wizard-of-Oz scenario
where linguistics
play a minor role
in the expression of emotion;
hence, text-based models
performed worse
and audio-based models
could draw
little additional information
from linguistics.

\section{Discussion \& Conclusion}
\label{sec:discussion}
In this section,
we jointly consider
all findings 
presented in the previous section
and connect them
to related work.

We begin with the most important
research question
that we 
set out to answer:
is recent progress on \gls{SER}
obtained by using newer
and bigger \gls{DL} models
monotonic?
Our answer was inconclusive.
Bigger and newer models
did not necessarily lead to 
better \gls{IID} or \gls{OOD} performance,
more robustness,
or more fairness.
In fact,
we also observed
opposing trends
with newer/bigger models
often showing lower fairness
or worse performance.
However,
one notable outcome from our work
is that any conclusions regarding progress
are pre-conditioned 
on the particular set of models
that are evaluated.
As shown in \cref{sec:audio},
varying the set of models included in the analysis
can have dramatic effects on observed correlation
between model performance and year or computational complexity.
This has major implications for any attempts to gauge progress
on a long-term horizon.
Replication studies 
like ours
are inevitably limited
by the effort
we may put in them
and our subjective biases
when selecting the methods that are included.
Seen in a different light,
reporting \glspl{CI}
on measures of progress
is vital
to avoid a misrepresentation
of the underlying findings.

The most important caveat underlying our work
is the strong impact of hyperparameters
on model performance,
as showcased in our \emph{tuning} phase.
The large variability
obtained for the models
which were optimised
during the tuning phase
indicates that
it is hard to identify
a clear winner
unless a sufficient amount of variations
are tested.
This point is further discussed
as a limitation of our work
in the next section.
However,
we note that similar computational constraints
plague most recent \gls{AI} works,
highlighting, once again,
the need
for exploring
a space of hyperparameters
as wide as possible
-- and certainly giving compared models
a `fair chance'
by considering at least a few different hyperparameters
and reporting the range of results.

Moreover, 
our work
can be positioned within the context
of the \emph{scaling hypothesis},
which is currently
driving the debate
in the broader \gls{AI} community~\citep{Kaplan20-SLF}
\footnote{Note that this thread of research is commonly referred to as ``neural scaling laws''. Given that these ``laws'' are merely experimental observations, rather than mathematical certainties, we prefer the softer term ``hypothesis''.}.
Is the solution to the \gls{SER} problem
merely a matter of scaling
to bigger models and bigger data?
We have shown that bigger models
do not necessarily lead to better performance,
though this was only done
when comparing \emph{across} architectures;
the scaling hypothesis is typically applied
within the same architecture family
(add more and wider layers).
Our computational resources
did not allow us to investigate this question
which is certainly worth pursuing
in future work.
However,
understanding the inner workings
of more successful models
can also help pave future advances
independent of scaling.
For instance, our probing analysis
revealed that state-of-the-art
transformer models
still struggle
with predicting
pitch variability.
Designing novel architectures
that are able to do so,
or pre-training tasks
that lead to
models doing so,
might be another
interesting avenue
for future research,
especially in the face
of scarce computational resources.

Ultimately,
the scaling hypothesis
is an important 
and pressing question
for our field
that our work
has not conclusively answered.
While our results
point against it,
we have not nearly
reached the levels of scale
commonly achieved
by ``frontier models'' 
(on the scale of GPT-4o).
In the case of academic research,
this amount of compute resources
is potentially unattainable
in the near-term future.
At the very least,
the recent stabilisation
of {\podcast},
and the existence
of streamlined training pipelines,
like our \emph{autrainer} toolkit~\citep{Rampp24-AAM},
represent an opportunity
to establish
a common benchmark
for naturalistic \gls{SER}
that will allow
comparisons
of long-term progress.

\section{Limitations}
\label{sec:limitations}

We are aware that our work suffers from certain limitations, which we aim to discuss here.
\begin{itemize}[align=left]
    \item[\emph{Insufficient data --}] From the multitude of available \gls{SER} datasets, we chose to focus on only two, with agreement in the ranking of model performance across datasets being moderate-to-high.
    See \cref{app:data} for a longer discussion on why we chose these two particular datasets.
    \item[\emph{Insufficient models --}] With hundreds of \gls{DL} papers published in the domain of \gls{SER} in recent years, it was impossible to replicate all of them.
    We opted to focus on the ones that were easier to replicate from a coding perspective.
    Our most important omission was the non-inclusion of large audio-language models~\citep{Triantafyllopoulos24-CAF, Latif25-CLL}, {\ie} models which connect an audio encoder to an \gls{LLM} decoder for joint audio-text modelling.
    While extremely promising, there were not many models available at the time when we conducted our experiments.
    \item[\emph{Insufficient tuning --}] Our investigation of hyperparameter tuning revealed that model performance heavily depends on the choice of parameters.
    Scaling up that search is contingent upon the availability of compute power.
    It could very well be that our choice of hyperparameters for the exploration phase may have favoured some models over others.
    \item[\emph{Inconclusive outcomes --}] The dependence of our results on the particular selection of models and the large impact of hyperparameters highlights the uncertainty of our findings.
    Nevertheless, counter to overall consensus, we did not observe a clear and consistent trend favouring bigger and more recent models.
    Furthermore, any observed gains can be attributed to a selection of hyperparameters that favoured some models more than others.
\end{itemize}
We acknowledge all of the above limitations.
They are all known problems that plague the entire \gls{AI} community.
In fact, our work can be seen in direct connection to works highlighting issues of \emph{comparability} and \emph{replicability of findings}~\citep{Snoek18-WCO, Lipton19-TTI, Choi19-OEC, Bouthillier19-URI}.
Our biggest takeaway is to interpret recent findings with the same care; hyperparameter tuning, rather than architectural innovation, may be responsible for reported empirical gains.
In particular, we urge for caution with regards to the \emph{scaling hypothesis}, {\ie} the hypothesis that bigger models trained on more data leads to better results.
The same limitations that plague our work are also important for works that provide confirmatory evidence for it -- perhaps even more so given the possibility of publication bias in favour of positive results.
As confirmation or refutation of this hypothesis stands to be a defining model for our field, we hope that our investigation serves as a reminder that the replication of \emph{findings} is oftentimes just as important as the replication of \emph{results}.

\section{Future work}
\label{sec:future}
We end our contribution
by considering potential pathways
for future research.
A major bottleneck
in such large-scale
replication studies
is the lack of compute capacity.
For academic institutions
with modest computational resources
this constraint can only be overcome
by intense cross-institutional collaboration;
resources can be pulled to increase the number
of investigated models and hyperparameters.
Shared authorship for shared compute
could be a model that facilitates
such large-scale collaboration;
perhaps we could envision
scientific publications
with longer lists of authors
as those seen
in other fields,
like physics.

Beyond that,
it is necessary to invest more effort
in interpreting
the ways in which 
models make their decisions (interpretability/explainability)
and the ways in which they fail (error analysis/fairness/robustness).
These analyses
can uncover 
the weak-points
of contemporary architectures
and inspire novel ways to mitigate them.
For instance,
the fact that transformer models
seem to be worse
at predicting pitch variations
than pitch itself
is a clue worth further investigation.

Finally, 
it is perhaps time
to begin considering
what may lie beyond
deep learning~\citep{Triantafyllopoulos24-BDL}.
Perhaps
we have reaped the benefits
of incorporating feature extraction
with classification
and learning both from data
(the so-called \emph{end-to-end} paradigm),
and further increasing the complexity of models
and the data they ingest
can only offer
diminishing returns.
Recent progress 
on \emph{reasoning} models
offers a promising avenue
to integrate pre-trained `world knowledge'
with contextual features analysis
to help overcome
the inherent uncertainty 
in predicting human emotions
in a naturalistic environment.

\section*{Acknowledgments}
This work has received funding from the DFG's Reinhart Koselleck project No.\ 442218748 (AUDI0NOMOUS).

\section{\refname}
\printbibliography[heading=none] 

\appendix

\subsection{Dataset selection}
\label{app:data}

The selection process that resulted in the two datasets used for benchmarking, {\aibo} and {\podcast}, was far from trivial.
In the present section,
we provide a justification for our selection criteria.

Our first point of consideration was the \emph{context} in which the portrayed emotions arose.
Ultimately, \gls{SER} systems that are deployed in real-world applications will inevitably have to process spontaneous human 
expressions that correspond to emotional triggers in the real world.
Improving performance on acted data does not necessarily equate an equal improvement on naturalistic data.  
This led us to exclude most commonly-used datasets such as {\iemocap}~\citep{Busso08-IIE}, \emph{RAVDESS}~\citep{Livingstone18-TRA}, \emph{CREMA-D}~\citep{Cao14-CCE}, or {\emodb}~\citep{Burkhardt05-ADO} for our in-domain training as they all comprise exclusively acted content\footnote{Nevertheless, we do use two of those datasets, {\iemocap} and {\emodb}, to gauge out-of-domain generalisation given the relative lack of naturalistic datasets.}.

We note that neither {\aibo} nor {\podcast} are perfect in this regard.
{\aibo} was collected in a very artificial Wizard-of-Oz scenario and the (underage) subjects were aware that they were being recorded.
As such, their portrayal of emotion might have been mediated by this `observer effect'.
However, using a Wizard is an established procedure in elicitation studies, with the underlying premise being that the elicited reactions are similar to the ones found in ``in-the-wild'' scenarios.
On the other hand, {\podcast} is also not free from criticism in this regard, given that its material was sourced from podcasts which aired in public domains.
Even though the subjects were (probably) not under the impression of being monitored, they were certainly aware that they were speaking in front of an audience.
While this is a valid context in which to measure human emotions, it is \emph{a} context nonetheless.

Eventually, we had to surrender to the fact that finding a dataset which encompasses the whole palette of human experience, with participants further being entirely unaware of their ever being monitored, would be impossible without resorting to unsolicited surveillance.
For this reason, we settled for {\aibo} and {\podcast} -- both encompassing only a specific snapshot of the human experience, which, limited though it may be, is still much wider than most datasets comprising solely acted emotional speech.

Our next desideratum was the \emph{coding scheme}, where we distinguish between schemes driven by \emph{theory}, and schemes driven by \emph{observation}.
The vast majority of \gls{SER} datasets follow a prescriptive, top-down approach and use a set of pre-defined emotional categories.
Usually, these correspond to -- or at the very least are inspired from -- Ekman's ``big-6''~\citep{Ekman99-BEM}, plus a ``neutral'' category to designate a complete lack of emotion.
This process is inherently \emph{model-based}.
In other words, it enforces a specific, pre-defined model of human emotions on the annotation process.
{\podcast} follows this recipe by adopting 8 categorical labels (plus 1 ``other'' class), which we further restrict to the four categories most commonly-found in \gls{SER} literature \{anger, happiness, neutral, sadness\}.

While this enables a seamless translation of its annotations to many other \gls{SER} corpora, it is still enforcing a particular model of human emotions and their expressions on the annotators.
This model is not without criticism and might be too limited to capture the entire spectrum of human experience~\citep{Barrett17-HEA}.

On the other end of the theory-observation divide are \emph{data-driven} approaches.
This is the paradigm that {\aibo} primarily adheres to. 
The set of candidate labels was developed iteratively by experienced annotators following a consensus approach.
Then, 
the data was annotated using an original set of 11 labels on the word-level.
Subsequently, a majority vote for the five annotators was taken over all word-level annotations in an utterance to derive an utterance-level annotation. 
The amount of labels was then reduced to 5 and 2 emotions, respectively, using a model-based mapping.
As such, the dataset comprises a set of very unique emotion labels which do not easily translate to other contexts -- but which, importantly, are tailored to the specific context that this dataset encapsulates.

The choice of coding scheme has important repercussions about the applicability of trained models and leads to a much larger debate.
In colloquial terms, our overarching goal is to build algorithms that are able to \emph{correctly classify} human emotions in \emph{all} contexts and in \emph{all} situations.
Following generic, model-based approaches eases the translation of model predictions to new contexts, but to the detriment of specificity.
On the other hand, building models that are tailor-made to specific scenarios suffers from scalability and risks the fragmentation of research efforts.
The search for ways to accommodate these two paradigms is a very important research direction, but one that was beyond the scope of our work.
Settling for one model-based and one data-driven dataset was a good compromise.

Finally, \emph{dataset size} and \emph{fame} were two last considerations.
Given our focus on \glspl{DNN}, we naturally favoured datasets with more training instances.
Moreover, we aimed for datasets that are more widely-used by, or at least known to, the broader \gls{SER} community.
Among the ones that are not acted, {\podcast} and {\aibo} were two reasonable choices.

\subsection{Models}
\label{app:models}

\renewcommand{\UrlFont}{\ttfamily\small}
This section
contains the details
for all pre-trained models
used in our work.

\cref{tab:app:models:os}
records
the configuration files
used to extract {\opensmile} features.
We accommodate both static and dynamic versions of the official feature sets.
For the \emph{dynamic} features,
we used a $2$-layered \gls{LSTM} model with $32$ hidden units, followed by mean pooling over time, one hidden linear layer with $32$ neurons and ReLU acivation, and one output linear layer; all hidden layers are followed by a dropout of $0.5$; these models are denoted with $-d-lstm$.
Additionally, we train $3$-layered \glspl{MLP} with $64$ hidden units each, a dropout of $0.5$, and ReLU activation for the \emph{static} features; these models are denoted with $-s-mlp$.

\cref{tab:app:models:crnn}
contains the paths 
to the model definitions
of \glspl{CRNN} models.
We used two particular instantiations introduced by \citet{Tzirakis18-ETE} (\emph{CRNN}$^{18}$) and \citet{Zhao19-SER} (\emph{CRNN}$^{19}$).

\cref{tab:app:models:cv}
contains the paths
to model code and pre-trained states
used for spectrogram-based models
that have been pre-trained on vision data.
We used \emph{AlexNet}, \emph{Resnet50}, all versions of \emph{VGG} ($^{11,13,16,19}$), EfficientNet-B0 ({\effnet}), the tiny, small, base, and large versions of \emph{ConvNeXt} ($^{t,b,s,l}$), and the tiny, base, and small versions of the \emph{Swin} Transformer ($^{t,b,s}$).
In all cases, we use the best-performing model state on \emph{ImageNet} as available in the \emph{torchvision-v0.16.0} package.
As features, we always used the Mel-spectrograms generated for \cnn, {\ie} $64$ Mels with a window size of $32$\,ms and a hop size of $10$\,ms; the resulting matrices were then replicated over the three dimensions to generate the $3$-channel input that is required by models designed for computer vision tasks.

\cref{tab:app:models:supervised}
lists the models pre-trained on supervised audio tasks.
We used {\cnn} and \emph{AST} pre-trained on \emph{AudioSet},
as well as the \emph{ETDNN} model pre-trained for speaker identification~\citep{Desplanques20-EEC}.
Furthermore, we fine-tuned \emph{Whisper}, albeit only its three smallest available variants of  ($^{t,b,s}$) due to hardware constraints.

Finally,
\cref{tab:app:models:ssl}
documents the paths of \gls{SSL} models,
which are largely the same as the ones used in \citet{Wagner23-DOT}.
In this work, we used the pretrained states from the \emph{base} and \emph{large} variants of {\wav} and {\hubert} (\wbase, \wlarge, \hbase, \hlarge), a multilingual model trained on \emph{VoxPopuli}~\citep{Wang21-VAL} (\wvox), a \emph{`robust'} version of \wav{} trained on more data~\citep{Hsu21-RW2} (\wrobust), as well as the pruned version of that model further finetuned for dimensional \gls{SER} on \emph{MSP-Podcast} that was presented in \citet{Wagner23-DOT} (\wemo).
Similar to \citet{Wagner23-DOT}, we add an output $2-$layered \gls{MLP} which takes the pooled hidden embeddings of the last layer as input.


\begin{table*}[!t]
    \centering
    \caption{
    Feature URLs for \opensmile.
    }
    \label{tab:app:models:os}
\begin{tabularx}{\textwidth}{|l|X|}
\toprule
    Features &  URL \\
    \midrule
    \emph{IS09-s,d} & \url{https://github.com/audeering/opensmile/blob/v3.0.2/config/is09-13/IS09\_emotion.conf}  \\
    \emph{IS10-s,d} &  \url{https://github.com/audeering/opensmile/blob/v3.0.2/config/is09-13/IS10\_paraling\_compat.conf}  \\
    \emph{IS11-s,d} &  \url{https://github.com/audeering/opensmile/blob/v3.0.2/config/is09-13/IS11\_speaker\_state.conf}  \\
    \emph{IS12-s,d} & \url{https://github.com/audeering/opensmile/blob/v3.0.2/config/is09-13/IS12\_speaker\_trait\_compat.conf}\\
    \emph{IS13-s,d} & \url{https://github.com/audeering/opensmile/blob/v3.0.2/config/is09-13/IS13\_ComParE.conf} \\
    \emph{IS16-s,d} & \url{https://github.com/audeering/opensmile/blob/v3.0.2/config/compare16/ComParE\_2016.conf} \\
    \emph{eGeMAPS-s,d} & \url{https://github.com/audeering/opensmile/blob/v3.0.2/config/egemaps/v02/eGeMAPSv02.conf} \\
\bottomrule
\end{tabularx}
\end{table*}

\begin{table*}[!t]
    \centering
    \caption{
    Model URLs for end-to-end \glspl{CRNN}.
    }   
    \label{tab:app:models:crnn}
\begin{tabularx}{\textwidth}{|l|X|}
\toprule
          Model &  URL \\
\midrule
          \emph{CRNN}$^{18}$ & \url{https://github.com/end2you/end2you/blob/master/end2you/models/audio/emo18.py} \\
    \emph{CRNN}$^{19}$ & \url{https://github.com/end2you/end2you/blob/master/end2you/models/audio/zhao19.py} \\
\bottomrule
\end{tabularx}
\end{table*}

\begin{table*}[!t]
    \centering
    \caption{
    Model URLs for computer vision models applied on spectrograms.
    }
    \label{tab:app:models:cv}
\begin{tabularx}{\textwidth}{|l|X|}
\toprule
          Model &  URL \\
    \midrule
    \emph{AlexNet} & \url{https://pytorch.org/vision/0.16/models/alexnet.html}  \\
    \emph{Resnet50} & \url{https://pytorch.org/vision/0.16/models/generated/torchvision.models.resnet50.html\#torchvision.models.resnet50} \\
    \emph{VGG}$^{11}$ & \url{https://pytorch.org/vision/0.16/models/generated/torchvision.models.vgg11.html\#torchvision.models.vgg11}\\
    \emph{VGG}$^{13}$ & \url{https://pytorch.org/vision/0.16/models/generated/torchvision.models.vgg13.html\#torchvision.models.vgg13}\\
    \emph{VGG}$^{16}$ & \url{https://pytorch.org/vision/0.16/models/generated/torchvision.models.vgg16.html\#torchvision.models.vgg16} \\
    \emph{VGG}$^{19}$ & \url{https://pytorch.org/vision/0.16/models/generated/torchvision.models.vgg19.html\#torchvision.models.vgg19} \\

    \emph{ConvNeXt}$^t$ & \url{https://pytorch.org/vision/0.16/models/generated/torchvision.models.convnext\_tiny.html\#torchvision.models.convnext\_tiny} \\
    \emph{ConvNeXt}$^s$ & \url{https://pytorch.org/vision/0.16/models/generated/torchvision.models.convnext\_small.html\#torchvision.models.convnext\_small} \\
    \emph{ConvNeXt}$^b$  & \url{https://pytorch.org/vision/0.16/models/generated/torchvision.models.convnext\_base.html\#torchvision.models.convnext\_base}\\
    \emph{ConvNeXt}$^l$  & \url{https://pytorch.org/vision/0.16/models/generated/torchvision.models.convnext\_large.html\#torchvision.models.convnext\_large}\\

    \emph{Swin}$^t$ & \url{https://pytorch.org/vision/0.16/models/generated/torchvision.models.swin\_t.html\#torchvision.models.swin\_t} \\
    \emph{Swin}$^s$ & \url{https://pytorch.org/vision/0.16/models/generated/torchvision.models.swin\_s.html\#torchvision.models.swin\_s} \\
    \emph{Swin}$^b$ & \url{https://pytorch.org/vision/0.16/models/generated/torchvision.models.swin\_b.html\#torchvision.models.swin\_b} \\

    {\effnet} & \url{https://github.com/lukemelas/EfficientNet-PyTorch/blob/master/efficientnet\_pytorch/model.py\#L27} \\
\bottomrule
\end{tabularx}
\end{table*}

\begin{table*}[!t]
    \centering
    \caption{
    Model URLs for supervised audio models.
    \emph{Whisper} models found at \url{https://huggingface.co/}.
    }
    \label{tab:app:models:supervised}
\begin{tabularx}{\textwidth}{|l|X|}
\toprule
          Model &  URL \\
          \midrule
        \cnn & \url{https://github.com/qiuqiangkong/audioset_tagging_cnn/blob/master/pytorch/models.py} \\
        \emph{ETDNN} & \url{speechbrain/spkrec-ecapa-voxceleb} \\
    
    \emph{Whisper}$^t$ & \url{openai/whisper-tiny} \\
    \emph{Whisper}$^s$ & \url{openai/whisper-small} \\
    \emph{Whisper}$^b$ & \url{openai/whisper-base} \\
\emph{AST} & \url{MIT/ast-finetuned-audioset-10-10-0.4593}\\
\bottomrule
\end{tabularx}
\end{table*}

\begin{table*}[!t]
    \centering
    \caption{
    Model URLs for self-supervised speech models (\url{https://huggingface.co/}).
    }
    \label{tab:app:models:ssl}
\begin{tabularx}{\textwidth}{|l|X|}
\toprule
          Model &  URL \\
    
        \midrule
        \hbase & \url{facebook/hubert-base-ls960} \\
        \hlarge & \url{facebook/hubert-large-ll60k} \\
        
        \wbase  & \url{facebook/wav2vec2-base} \\
        \wlarge & \url{facebook/wav2vec2-large} \\
        \wvox & \url{facebook/wav2vec2-large-100k-voxpopuli} \\
        \wrobust & \url{facebook/wav2vec2-large-robust} \\
        \wemo & \url{audeering/wav2vec2-large-robust-12-ft-emotion-msp-dim} \\
\bottomrule
\end{tabularx}
\end{table*}

Our text-based models
are listed in \cref{tab:app:models:text}.
While we use the same acronym
to refer to both English
and German version
of some models,
these do not always correspond
to the same model state
as some models were trained
for a single language.
We only use the ``second-generation''
\glspl{LLM} for both languages
as they were trained on multilingual data.

\begin{table*}[t]
    \centering
    \caption{
    Text-based models
    used for \gls{SER}
    on {\aibo} and {\podcasteleven}.
    All models available at \url{https://huggingface.co/}.
    \scriptsize
    }
    \label{tab:app:models:text}
    \begin{tabularx}{\textwidth}{|l|X|}
        \toprule
        Model &  URL \\
        \midrule
        \multicolumn{2}{|c|}{\emph{English models}} \\
        \midrule
        {\bert}            & \url{google-bert/bert-base-uncased}\\
        \emph{RoBERTa}     & \url{FacebookAI/roberta-base}\\
        \emph{DistilBERT}  & \url{distilbert/distilbert-base-uncased}\\
        \emph{Electra}     & \url{google/electra-base-discriminator}\\
        \emph{Llama-2}     & \url{meta-llama/Llama-2-7b-chat-hf}\\
        \emph{Llama-3}     & \url{meta-llama/Meta-Llama-3-8B-Instruct}\\
        \emph{Mistral}     & \url{mistralai/Mistral-7B-Instruct-v0.2}\\
        \midrule
        \multicolumn{2}{|c|}{\emph{German models}} \\
        \midrule
        {\bert}            & \url{dbmdz/bert-base-german-uncased}\\
        \emph{RoBERTa}     & \url{T-Systems-onsite/german-roberta-sentence-transformer-v2}\\
        \emph{DistilBERT}  & \url{distilbert/distilbert-base-german-cased}\\
        \emph{Electra}     & \url{german-nlp-group/electra-base-german-uncased}\\
        \emph{Llama-2}     & \url{meta-llama/Llama-2-7b-chat-hf}\\
        \emph{Llama-3}     & \url{meta-llama/Meta-Llama-3-8B-Instruct}\\
        \emph{Mistral}     & \url{mistralai/Mistral-7B-Instruct-v0.2}\\
        \bottomrule
    \end{tabularx}
\end{table*}


\subsection{Codecs}
\label{app:codecs}
We investigated the robustness
of {\wser} on the following codecs:
\begin{itemize}[align=left]
    \item[\emph{MP3}] (MPEG-1/2 Layer-3) combines psychoacoustic modeling and compression of the modified discrete cosine transform~\citep{Brandenburg93-MAA}. We used the \emph{LAME} version available in \emph{ffmpeg}~\footnote{\url{https://lame.sourceforge.io/}}.
    Its psychoacoustic model was tuned on manual listening tests and focuses on maintaining the quality of specific frequency bands.
    \item[\emph{AAC}] (Advanced Audio Codec) was designed as a follow-up to \emph{MP3}~\citep{Brandenburg93-MAA}. It improves on it by -- among ohters -- increasing the frequency resolution.
    \item[\emph{Opus}] combines a short- and a long-term (linear) predictor. 
    \item[\emph{EnCodec}] is one of the most recent, data-driven codecs relying on \glspl{DNN}~\citep{Defossez22-HFN}.
    It is a simple, \gls{CNN}-based auto-encoder with an intermediate steps of residual vector quantisation.
    It has been trained on a large dataset comprising multiple audio sources.
    \item[\emph{SemantiCodec}] aims for higher compression rates by incorporating a very powerful encoder (\emph{AudioMAE}) and a more powerful \gls{DDPM} decoder~\citep{Liu24-SAU}.
    However, the improvement in compression rates comes with a substantial penalty on computational overhead.
\end{itemize}

\subsection{Robustness to additive noise}
\label{app:robustness}
\cref{tab:rob:cat} presents detailed robustness results for all models of our exploration phase.

\begin{table}[t]
    \centering
    \caption{
    \gls{UAR} on the original and noisy versions of the {\podcasteleven} test set.
    We use the models trained in the exploration phase of \cref{sec:audio}
    without additional retraining, {\ie} all models have only been trained on clean data.
    }
    \label{tab:rob:cat}
    \begin{tabular}{c|cc}

    \toprule
    \textbf{Model} & \textbf{Original} & \textbf{Noisy}\\
    \midrule
        \emph{IS09-s-mlp} & .250 &  .250\\ 
        \emph{IS09-d-lstm} & .469 &  .337\\  
        \emph{IS10-s-mlp} & .504 &  .373\\ 
        \emph{IS10-d-lstm} & .486 & .366\\ 
        \emph{IS11-s-mlp} & .502 &  .385\\ 
        \emph{IS11-d-lstm} & .499 &  .374\\ 
        \emph{IS12-s-mlp} & .501 &  .388\\ 
        \emph{IS12-d-lstm} & .496 &  .368\\ 
        \emph{AlexNet} & .250 &  .250\\ 
        \emph{IS13-s-mlp} & .503 &  .408\\ 
        \emph{IS13-d-lstm} & .505 &  .360\\
        \emph{eGeMAPS-s-mlp} & .479 &  .348\\ 
        \emph{eGeMAPS-d-lstm} & .491 &  .353\\ 
        \emph{Resnet50} & .510 &  .419\\  
        \emph{IS16-d-mlp} & .498 &  .353\\ 
        \emph{IS16-s-lstm} & .497 &  .387\\
        \emph{VGG}$^{11}$ & .459 &  .368\\ 
        \emph{VGG}$^{13}$ & .250 &  .250\\ 
        \emph{VGG}$^{16}$ & .471 &  .402\\ 
        \emph{VGG}$^{19}$ & .250 &  .250\\ 
        \emph{CRNN}$^{18}$ & .514 &  .419\\    
        \effnet & .537 &  .432\\ 
        \emph{CRNN}$^{19}$ & .503 &  .406\\ 
        \wbase & .250 &  .250\\
        \wlarge & .250 &  .250\\ 
        \emph{ConvNeXt}$^t$ & .518 &  .435\\ 
        \emph{ConvNeXt}$^s$ & .531 &  .449\\ 
        \emph{ConvNeXt}$^b$ & .513 &  .427\\
        \emph{ConvNeXt}$^l$ & .518 &  .423\\
        \emph{ETDNN} & .528 &  .450\\ 
        \cnn & .524 &  .461\\ 
        \hbase & .250 &  .250\\ 
        \hlarge & .570 &  .519\\ 
        \emph{Swin}$^t$ & .269 &  .244\\ 
        \emph{Swin}$^s$ & .250 &  .250\\ 
        \emph{Swin}$^b$ & .250 &  .250\\ 
        \emph{AST} & .396 &  .323\\  
        \wrobust & .555 &  .430\\ 
        \wvox & .561 &  .431\\ 
        \wemo & .609 &  .546\\ 
        \emph{Whisper}$^t$ & .387 &  .306\\
        \emph{Whisper}$^s$ & .405 &  .302\\ 
        \emph{Whisper}$^b$ & .250 &  .250\\ 
    \bottomrule
    \end{tabular}
\end{table}

\subsection{Linguistic content}
\label{app:linguistics}

\begin{figure*}
    \includegraphics[width=.49\textwidth]{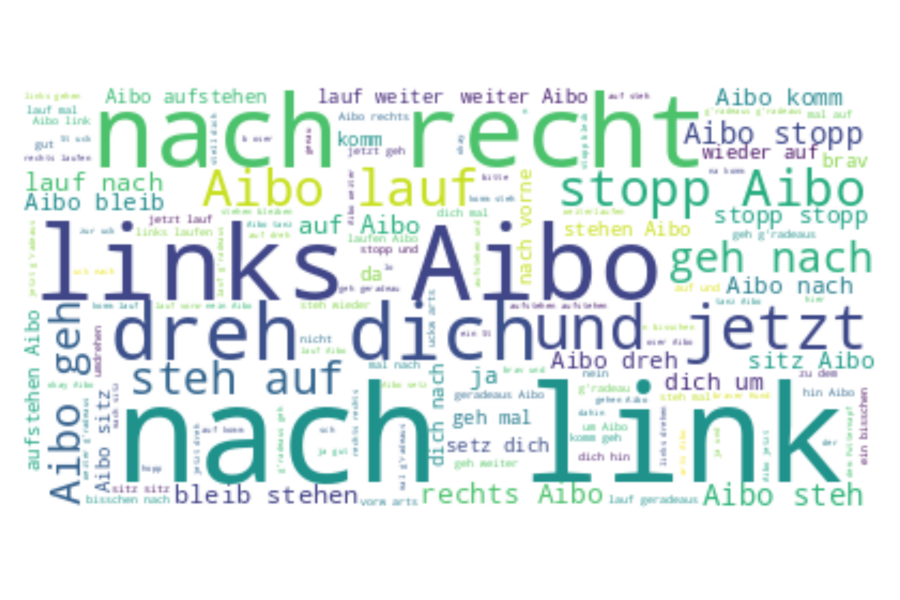}~%
    \includegraphics[width=.49\textwidth]{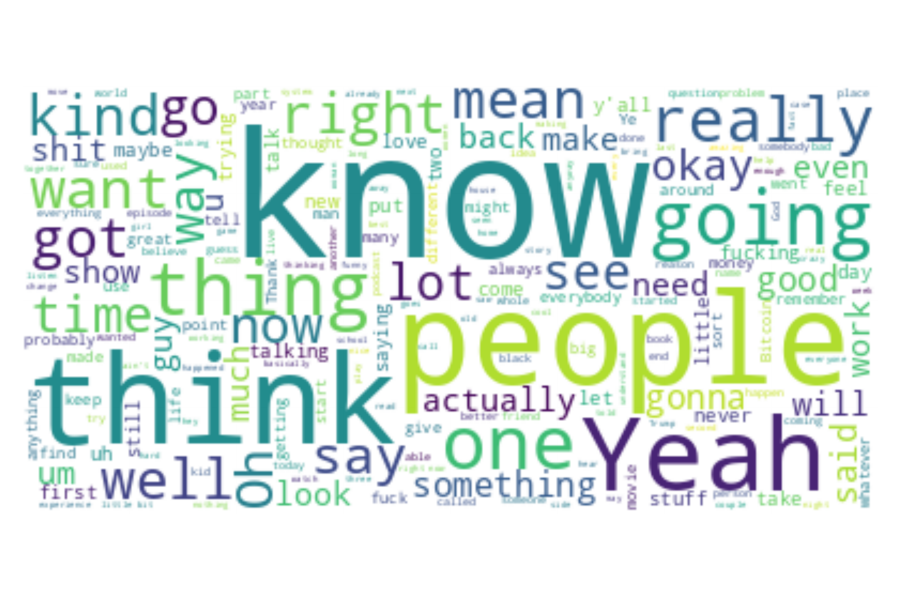}
    \caption{
    Wordclouds visualising bi-gram occurrence frequency for {\aibo} (left) and {\podcast} (right).
    }
    \label{fig:wordcloud}
\end{figure*}

\cref{fig:wordcloud}
shows the word distribution
for {\aibo} and {\podcast}\footnote{Computed using the default parameters of the Python \emph{WordCloud} package (\url{https://amueller.github.io/word_cloud/}).}.
We observe substantially higher variability
for {\podcast},
which is consistent
with its in-the-wild procurement.
{\aibo},
on the other hand,
comprises a much more constricted,
intent-oriented vocabulary,
featuring mostly
simply commands (``geh nach links''),
as well as
the frequent repetition
of the word \emph{Aibo}.

\end{document}